\documentclass[%
 aip,
 jcp,%
 amsmath,amssymb,
 preprint,%
]{revtex4-1}
\usepackage{placeins}
\usepackage[]{placeins}
\usepackage{graphicx, subfigure}
\usepackage{layout}
\usepackage{dcolumn}
\usepackage{bm}
\usepackage{booktabs,multirow}
\usepackage{hhline}
\usepackage{color}

\begin{document}
\title{Deposition of model chains on surfaces: anomalous relation between flux and stability }


\author{Pritam Kumar Jana}
\email{pjana\_01@uni-muenster.de}

  \affiliation{Westf\"{a}lische Wilhelms-Universit\"{a}t M\"{u}nster, Institut f\"{u}r Physikalische Chemie,
  Corrensstr. 28/30, 48149 M\"{u}nster, Germany}

\author{Andreas Heuer}%
\email{andheuer@uni-muenster.de}
\affiliation{
Westf\"{a}lische Wilhelms-Universit\"{a}t M\"{u}nster, Institut f\"{u}r Physikalische Chemie,
  Corrensstr. 28/30, 48149 M\"{u}nster, Germany
}%

\date{\today}

\begin{abstract}

Model chains are studied via Monte Carlo simulations which are
deposited with a fixed flux on a substrate. They may represent, e.g.,  stiff
lipophilic chains with an head group and tail groups mimicking the 
alkyl chain. After some subsequent fixed simulation time we determine the final
energy as a function of flux and temperature. Surprisingly, in some range of
temperature and flux the final energy increases with decreasing
flux. The physical origin of this counterintuitive observation is
elucidated. In contrast, when performing equivalent cooling
experiments no such anomaly is observed. Furthermore, it is elaborated whether flux experiments give rise to configurations with lower energies as compared to cooling experiments. These results are related
to recent experiments by the Ediger group where very stable configurations of glass-forming systems have been generated via flux experiments.
\end{abstract}

 \maketitle
\section{\label{sec:introduction}Introduction}

The growth behavior of molecules, adsorbed on surfaces, has been studied extensively \cite{Krug, Wang,
Venables, Haas, Vardavas, Nurminen, Sabiryanov}. This class of experiments can be guided by different key questions. (1)
Using prepatterned surfaces the adsorbed molecules may adopt the same prepatterned structure. In this way one can tune the structure
formation of molecules with interesting functions \cite{Wang, zhong, Briseno,Fuchs}. The ability to inherit the underlying
structure to the adsorbed molecules has been also analysed from a theoretical perspective; see, Refs. e.g. \cite{Kalischewski_1,
Kalischewski_2, Lied}
(2) From practical perspective one may optimize the properties of these thin film devices. This may be relevant for 
field effect transistors \cite{tsumura, Burroughes_1, assadi, paloheimo,
Horowitz}, organic light-emitting diodes  \cite {Burroughes_2, YutakaOhmori, braun}, or, more generally, opto-electronics
\cite {Burroughes_2, Yu, Sirringhaus, Yang}. (3) Using anisotropic molecules one may want to generate directional order of
the molecules on the surface. This order effect, already present for molecules with a large
aspect ratio \cite{Hopp,bellier-castella,Palermo}, become particularly pronounced for oligomers or polymers. Electronic
properties of organic molecular semiconductors can be finely tuned by modifying the chemical structure of their
constituting molecules \cite{Garnier}. Recently it has been shown that also
lipophilic alkane-chains with nucleobases as the respective headgroups display highly ordered structures.
This specific systems may also be used as the basis of electronics \cite{Bai}. (4) By the adsorption of the molecules on a
substrate, which was cooled below the glass transition temperature of this system, Ediger and coworkers managed to generate
glassy films with an enthalpy which is lower than the enthalpy obtained after cooling  a bulk sample in its glassy
state \cite{Ediger}. It was shown that in agreement with expectation the resulting enthalpy was correlated with the
applied flux: the smaller the flux the lower the resulting enthalpy.
One may generally ask, whether for all systems the enthalpy monotonously depends on the chosen flux. 

Inspired by the major
interest in the ordering process of anisotropic molecules on surface we specifically discuss this question for stiff
oligomers, containing a headgroup and a small tail. For a theoretical analysis of molecular systems on surfaces one often
resorts to lattice-gas models due to its simplicity; see, Refs. e.g., \cite {Pastor_1, Pastor_2, Rzysko}.
As a minimum system we consider straight trimers on a quadratic lattice.

The general setup of the simulation is shown in Fig.\ref {Fig:
schemetic diagram}. We start with a time period during which $N$
chains are deposited on a surface with constant flux at some fixed
temperature $T$. In the subsequent evolution period $t_{sim}$ the
system evolves further without the adsorption of additional
molecules. At the end of the simulation we record the potential
energy as a measure of the degree of equilibration. At very low
temperatures one may expect that the system cannot reach its
equilibrium state during $t_{sim}$. Intuitively, one might expect
that for high flux and low temperature the chains form some
disordered high-energy configuration during the first time
interval which are stabilized by the immediate arrival of new
chains. So, the general expectation is that the final state has a higher energy for higher initial
flux, though there  will be some aging during the evolution period $t_{sim}$.
 In contrast, for small flux unfavorable configurations have
sufficient time to dissolve so that the final configuration might
typically correspond to a low-energy structure.

\begin{figure}
  \centering
   \includegraphics[width=0.45\textwidth]{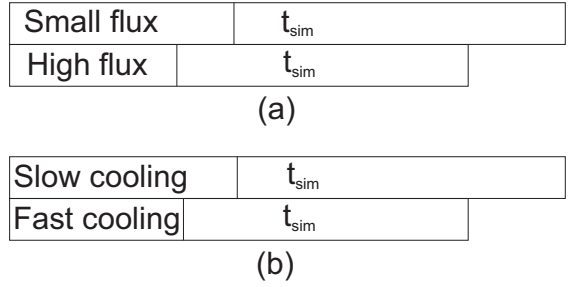} 

  \caption{\label {Fig: schemetic diagram} \small Schematic presentation of the set-up of the different simulations.}

\end{figure}

For comparison we also perform simulations where the initial deposition protocol is substituted by a cooling protocol as
also sketched in Fig.\ref {Fig: schemetic diagram}. Starting with an equilibrium configuration of all $N$ chains at a high
temperature we study the final energy in dependence on the cooling rate. Here the analogous argument should hold: the slower
the schedule of cooling down, the lower the resulting energy \cite{Pablo}. This effect is the basis of the well-known
simulated annealing technique to find low-energy states \cite{Kirkpatrick1983}.

Here we show that for the chosen model system the flux simulations
do not fulfill the general expectation as formulated above. Rather
for some temperature regime we find a surprising effect that
upon decreasing flux the resulting potential energy increases.
Furthermore we show that apart from this specific parameter regime
flux simulations  seem to be more efficient to find low-energy
configurations as compared to cooling simulations. The paper is
organized as follows: in Sec. \ref{sec:model} and Sec.
\ref{sec:simulationdetails} we describe the model and simulation
details, respectively, whereas in Sec. \ref{sec:results and
analysis} the results and the explanation of our observation for
the two different approaches to the final system parameters N
(number of chains) and T (temperature) are presented, i.e. either
by successive adsorption of chains at fixed temperature or by a
temperature decrease at fixed number of chains. Finally, we
summarize in Sec. \ref{sec:summary}.

\section{\label{sec:model}Model}

\begin{figure}
  \centering
   \includegraphics[width=0.45\textwidth]{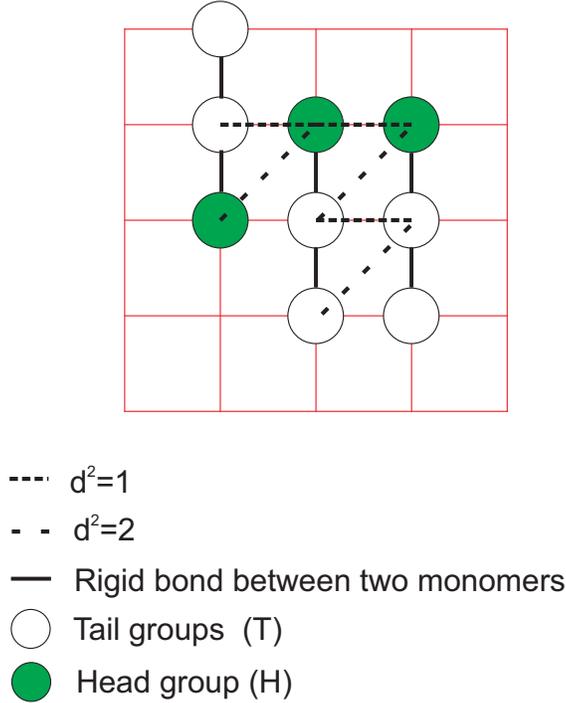} 
  \caption{\label {Fig: Modelsystem} \small The model system, consisting of trimers.}
\end{figure}

\begin{figure}
  \begin{center}
     \includegraphics[width=0.25\textwidth]{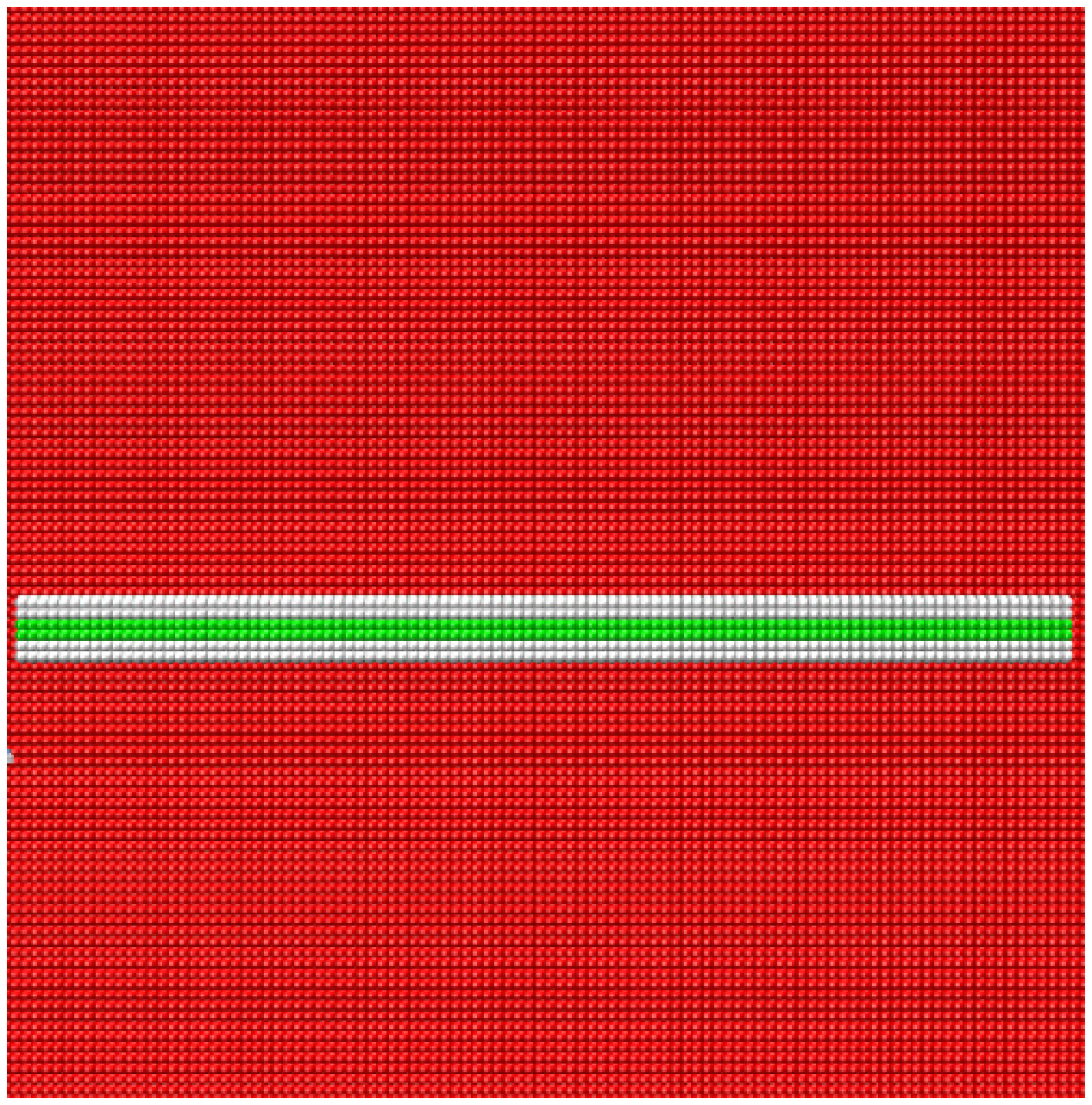}
  \end{center}

  \begin{center}
    (a)
  \end{center}

  \begin{center}
     \includegraphics[width=0.25\textwidth]{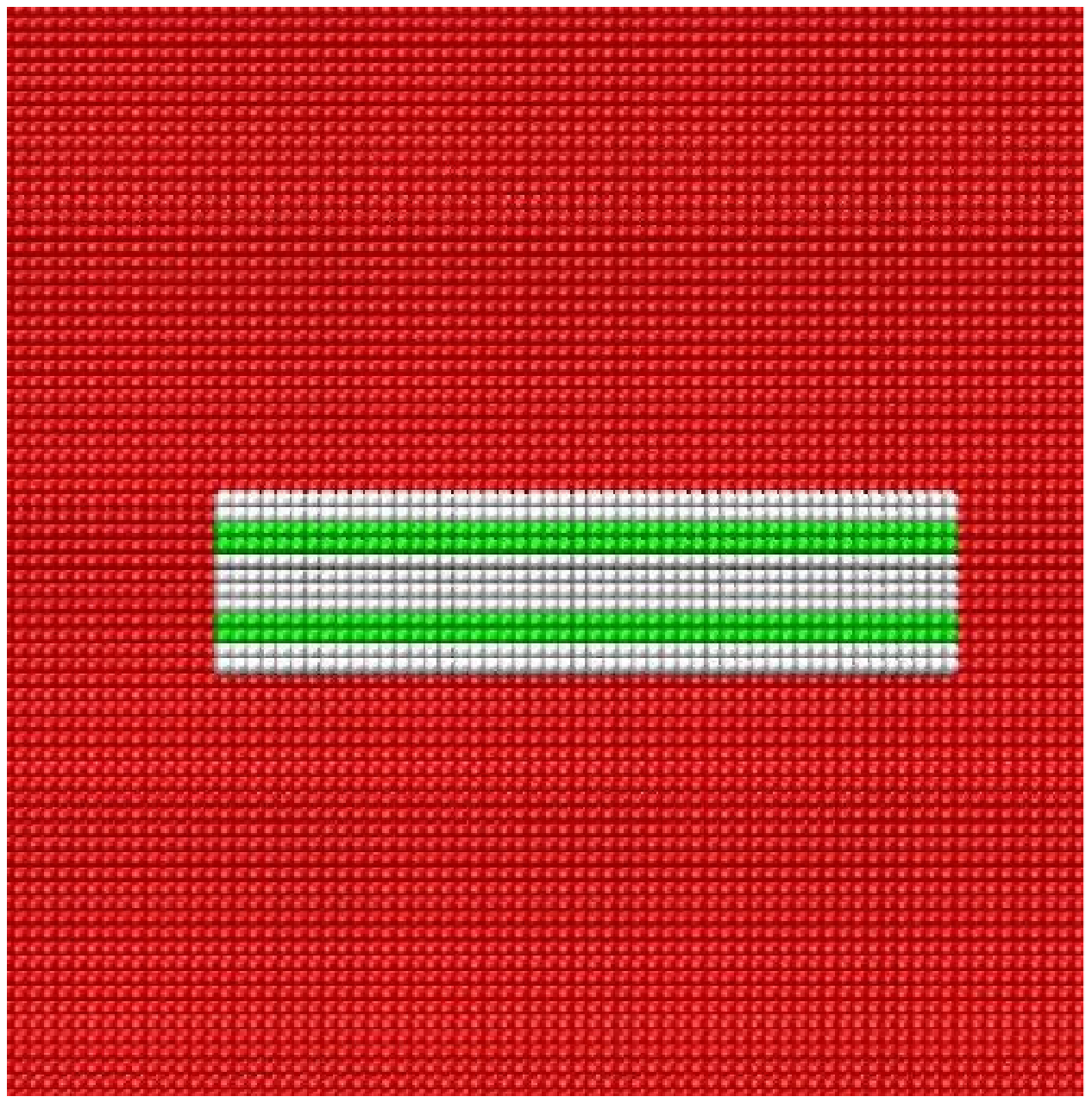}
  \end{center}

  \begin{center}
    (b)
  \end{center}

  \begin{center}
     \includegraphics[width=0.25\textwidth]{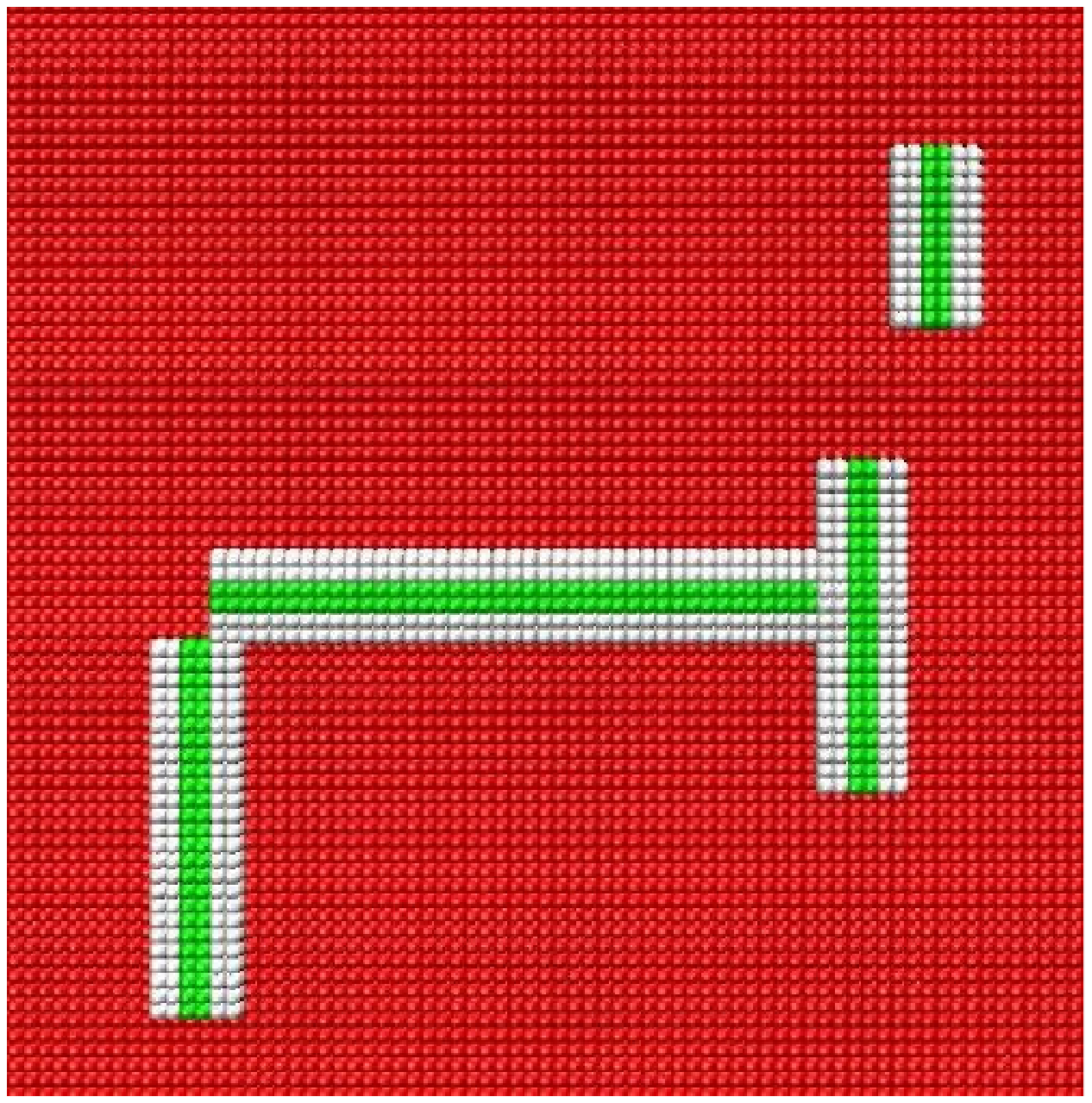}
  \end{center}

  \begin{center}
    (c)
  \end{center}

  \caption{\label {Fig: Struc} \small (a) The ordered structure (O1), (b) The ordered structure O2, and
                                 (c) The disordered structures (DO). Different arrangement of the rod like linear trimer HTT.}

\end{figure}

In our model system we consider N=200 trimers on a square lattice of length $L=120$ with periodic boundary conditions.
Each trimer consists of one head group (H) and two tail groups (T) and is chosen to be rigid. Each monomer occupies one
site of the lattice and the trimers can be arranged along both axes. Naturally, two chains cannot cross each other.
The six different types of nearest-neighbor interaction energies
between monomers belonging to different molecules are denoted $u_{ij_1}$, $u_{ij_2}$
with $i,j \in \{H,T\}$.  The index '1' and '2' indicates the squared distance between the respective monomers;
see Fig. \ref{Fig: Modelsystem}. There is the largest interaction between head-groups (HH), in order to guarantee local clustering of them.
Furthermore, the tail groups of adjacent chains also attract each other (TT). For alkyl chains this interaction is of van der waals type.
The specific energies for the present simulation are listed in Tab. \ref{Tab: Energy}, given in dimensionless units. Later on it will become clear that 
the key results of this work do not depend on the specific choice of the interaction parameters as long as the head groups
are interacting most strongly and the tail groups have a weaker but still finite interaction.
\begin {table}
\begin{center}
  \renewcommand{\arraystretch}{1.2}
    \begin{tabular}{|c|c|c|}
      \hline
      \multirow{2}{*}{Pair} &\multicolumn{2}{c|}{Interaction energy}\\
      \hhline{~--}
      & d$^2$=1&d$^2$=2 \\\hline
      HH &  -1.00 & -0.10\\ \hline
      TT &  -0.20 & -0.10\\ \hline
      HT &   0.25 &  0.03\\ \hline

    \end{tabular}
  \end{center}
\caption{\label{Tab: Energy}\small Interaction energies of the model chains.}
\end{table}

For the subsequent discussion we define the  two ordered
structures O1- and O2- as displayed in Fig. \ref{Fig: Struc}, having
one or two stripes. The energies are -2.124 and  -2.206 per chain,
respectively. Structures with more parallel are hardly seen in our simulations. 
As compared to O1- the ordered structure O2- displays additional
interactions of the tail groups between both stripes at the
expense of one missing interaction of two parallel pairs of chains. Note
that the energy contribution of the interaction of the different
tail groups is proportional to the length of the structure. As a
consequence for smaller values of $N$ ($N \leqslant 36$) the structure O1-
is more favorable than O2. This generic property of the chain
model will become important for our later discussion.  In
particular for high fluxes and/or low temperatures disordered
arrangements will prevail, as shown, e.g., in Fig.  \ref{Fig:
Struc}. This specific structure has an energy of -2.082.

\section{\label{sec:simulationdetails}Simulation details}

Monte Carlo simulations, based on the Metropolis criterion,  have
been performed. In one simulation step all chains, present at that
time, are attempted to move. The move class contains a rotational
as well as a translational motion. For the rotational motion one
randomly selects one of the two end particles and rotates the
total chain by 90$^\circ$ around this particle. For the
translational motion the chain is shifted by $n$ lattice sites
($n\geq1$) in the x- or the y-direction. Here we distinguish
whether this motion is restricted to a nearest-neighbor site
($n=1$: local move) or to any site ($n\geq1$: global move). For
most of the results in this work we use global moves. However, we
show that the key results are the same for local moves.

As mentioned already in the Introduction the non-equilibrium
simulation period either involve a constant flux of particles or a
temperature reduction. In the first case one fixes the flux and
the temperature. The first time period stops when all $N$
particles are on the substrate. In second case one starts with an
equilibrated configuration at $T_h=1$ and reduces the
time-dependent temperature $T(t)$ from the initial temperature
$T_h$ to the final temperature $T_f$ via

\begin{equation}
  T(t)=T_h +\frac{t}{t_{cool}}\times(T_f-T_h).
\end{equation}
the cooling rate $r$ can be defined as $r = (T_h-T_f)/t_{cool}$.
For reasons, discussed below, this type of simulation will be only
performed for  $T_f = 0.19$.

For one simulation run the resulting potential energy $u$ is
defined as the average over the last 25\% of the simulation period
of length $t_{sim}$. Furthermore we repeat at least 50 and up to
400 independent simulations, depending on the required statistical
accuracy. If not mentioned otherwise, we choose
$t_{sim}=1\times10^7$ Monte Carlo steps.

\section{\label{sec:results and analysis}Results and analysis}
\subsection{Flux simulations}
\begin{figure}
  \centering
  \includegraphics[width=0.45\textwidth]{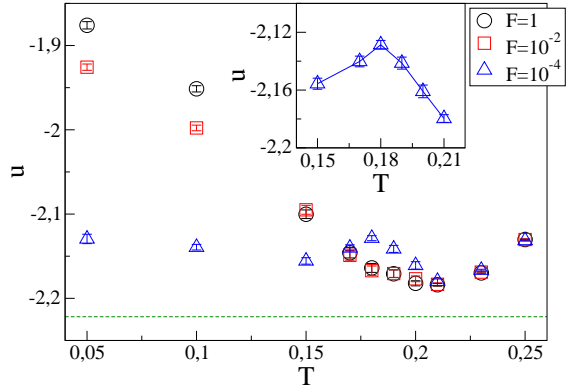}
  \caption{\label {Fig: E_T}\small Variation of $u$ with $T$ at different $F$.
                                   The dotted horizontal line indicates the minimum $u$ of the system.
                                   In the inset the data for F=10$^{-4}$ are highlighted. The solid
                                   line is a guide to the eyes.}
\end{figure}

\begin{figure}{
  \centering
  \includegraphics[width=0.45\textwidth]{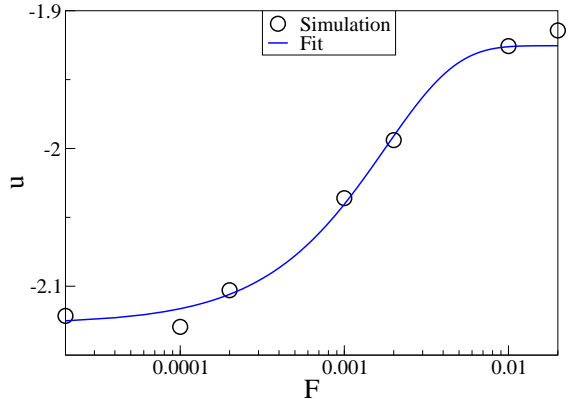}
  \caption{\label {Fig: E_F_05} \small Variation of $u$ with $F$ at temperature $T$=0.05 for global movement.
                                    The solid line is an exponential plot fit $u(F)=a+b\times exp(-F/F_0)$ with $F_0=1.8\times 10^{-3}$.}}
\end{figure}

First, we start with the flux simulations. The dependence of the potential energy  on flux and temperature, i.e. $u(F,T)$,
is shown in Fig. \ref {Fig: E_T}. In the limit of high
temperatures the time $t_{sim}$ is long enough to generate an
equilibrium structure, independent of the initial condition as determined by the chosen flux $F$. Indeed this independence can
be seen for the temperature range $T \ge 0.23$. Naturally, due to
the increasing relevance of entropic effects with increasing
temperature one observes an increase of $u$. In contrast, in the
limit of low temperatures the equilibration time by far exceeds
the chosen simulation time $t_{sim}$. Thus, the final energy still
reflects the situation directly after the flux period. For a high
flux the time interval between the successive deposition of
trimers is so short that by the advent of new chains any temporary
disordered structure may be stabilized, giving rise to high-energy
structures. Indeed, we see for $T=0.05$ that within error bars the
energy monotonously increases with increasing flux. Interestingly, a closer analysis of the flux-dependence reveals a simple exponential dependence with $F_0=1.8\times 10^{-3}$. 
See Fig. \ref{Fig: E_F_05}. A closer analysis shows that for $F << F_0$ the system is mainly trapped in the O1-structure whereas otherwise the system displays lot of disorder.

Following the arguments for low and high temperatures one would
thus expect that for some intermediate temperature a minimum
energy $u$ is observed which, furthermore, should increase with
increasing flux. As shown in Fig. \ref {Fig: E_T}, this simple
expectation is not fulfilled. First, the temperature dependence is
more complicated. As clearly seen for, e.g., $F = 10^{-4}$ there
exists a local maximum of $u$ for $T = 0.18$.

\begin{figure}
  \centering
  \includegraphics[width=0.45\textwidth]{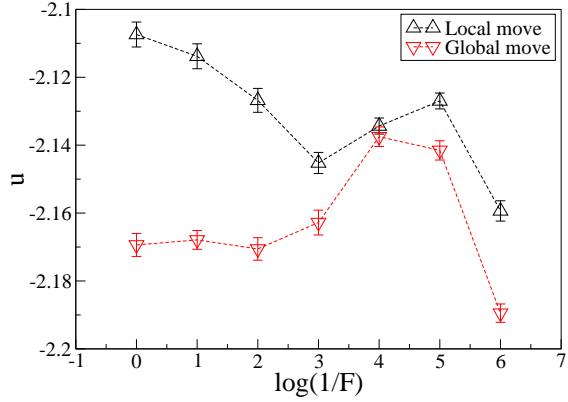}
  \caption{\label {Fig: E_F} \small Variation of $u$ with $F$ at temperature $T$=0.19 for global and local movement.
                                    The dashed line is a guide to the eyes.}
\end{figure}

\begin{figure}
  \centering
  \includegraphics[width=0.45\textwidth]{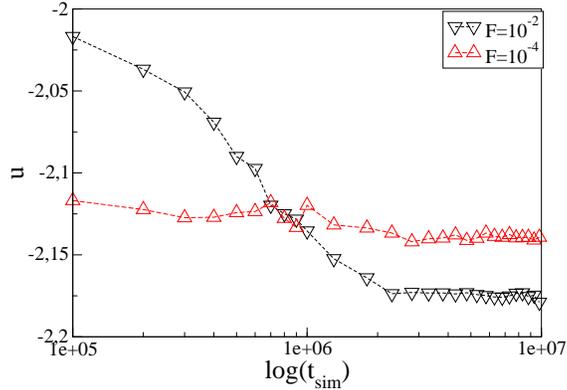}
  \caption{\label {Fig: E_tsim_F} \small Variation of $u$ with evolution of time for different values of the flux at temperature $T$=0.19.}
\end{figure}

\begin{figure}
  \begin{minipage}[b]{0.45\linewidth}
    \begin{center}

       \includegraphics[width=0.45\textwidth]{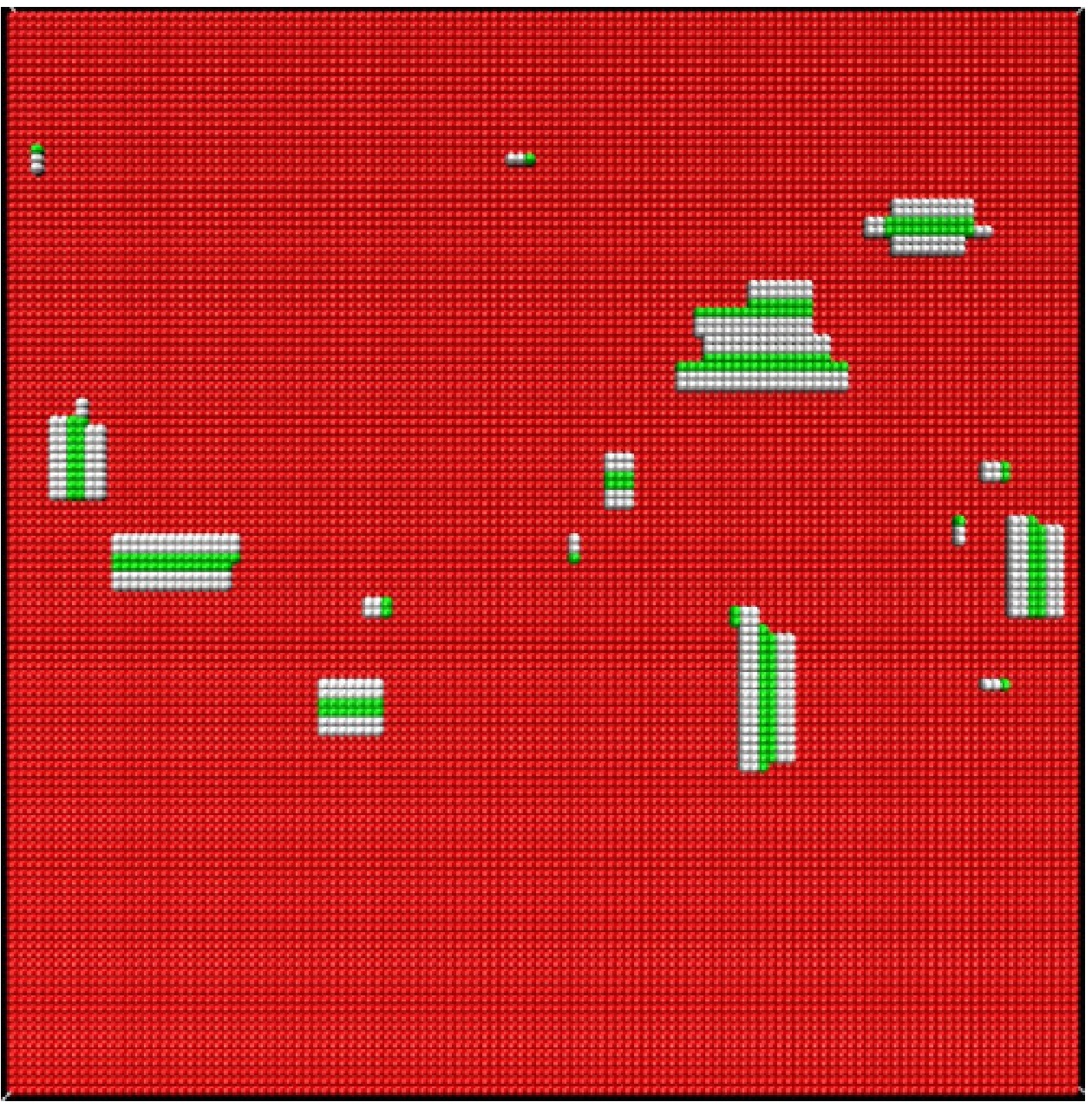}
    \end{center}

    \begin{center}
      (a)
    \end{center}

    \begin{center}

       \includegraphics[width=0.45\textwidth]{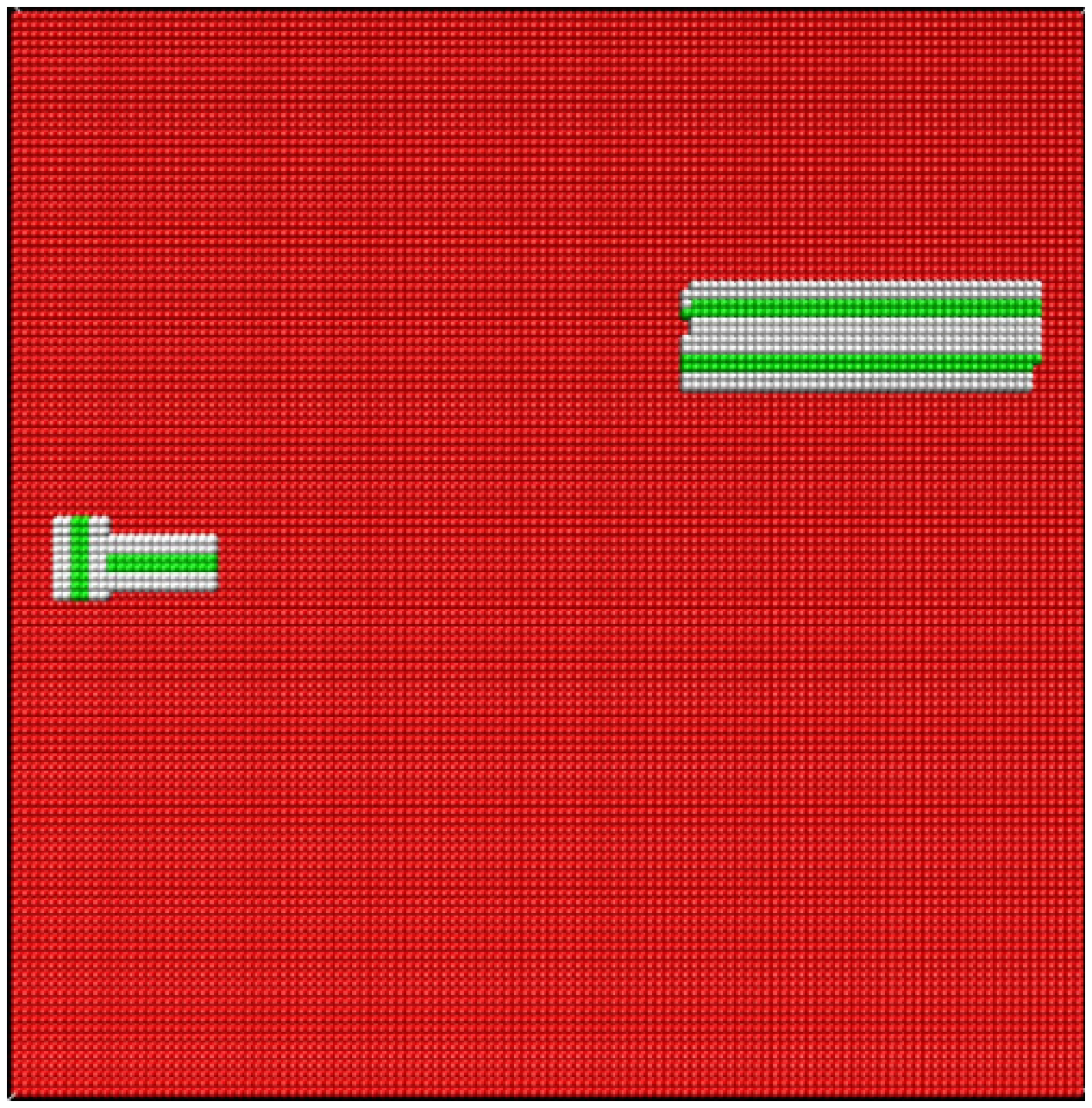}
    \end{center}

    \begin{center}
      (b)
    \end{center}

     \begin{center}
       \includegraphics[width=0.45\textwidth]{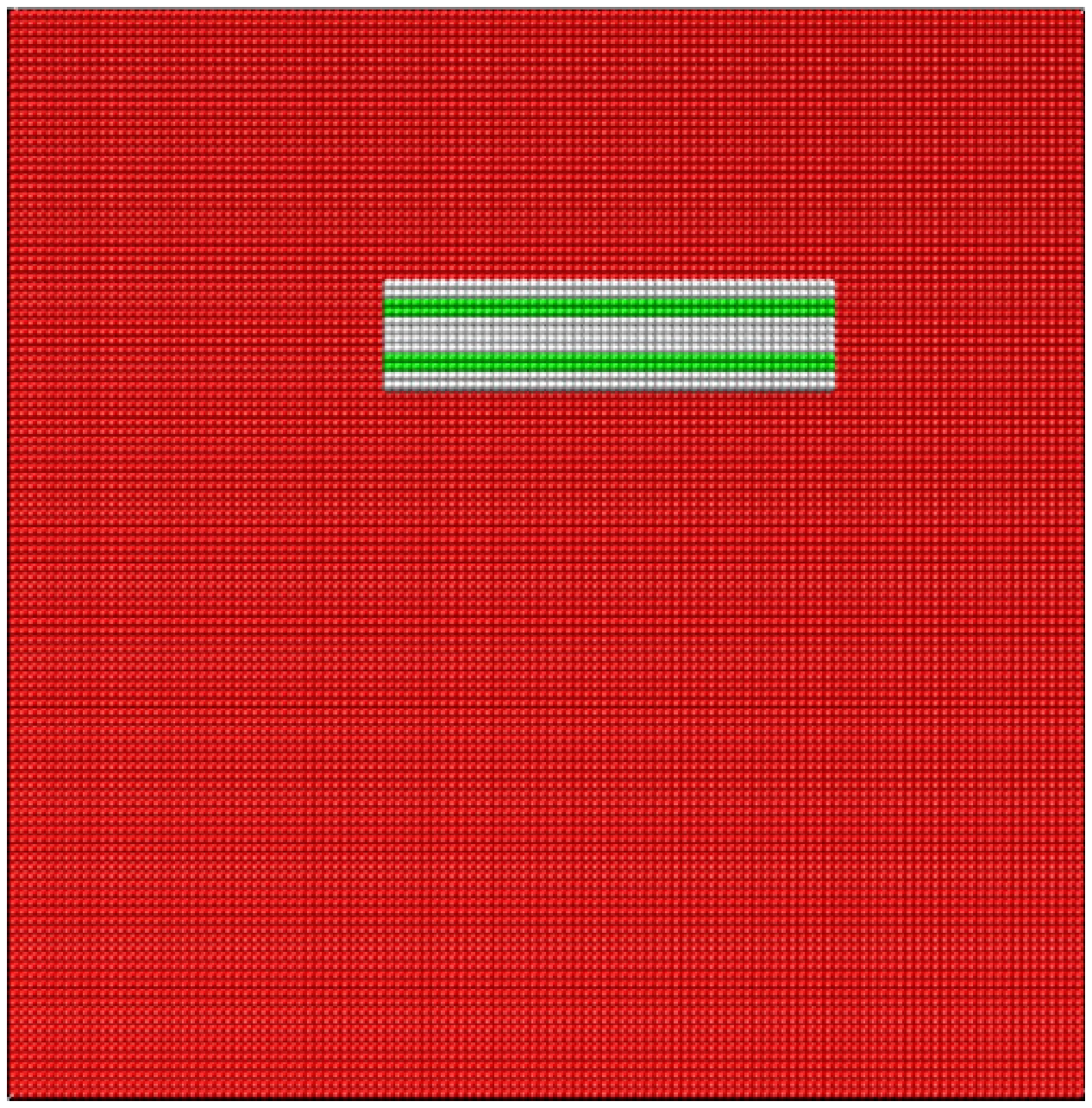}
    \end{center}

    \begin{center}
      (c)
    \end{center}

   \end{minipage}
   \hspace{0.5cm}
   \begin{minipage}[b]{0.45\linewidth}
     \begin{center}
          \includegraphics[width=0.45\textwidth]{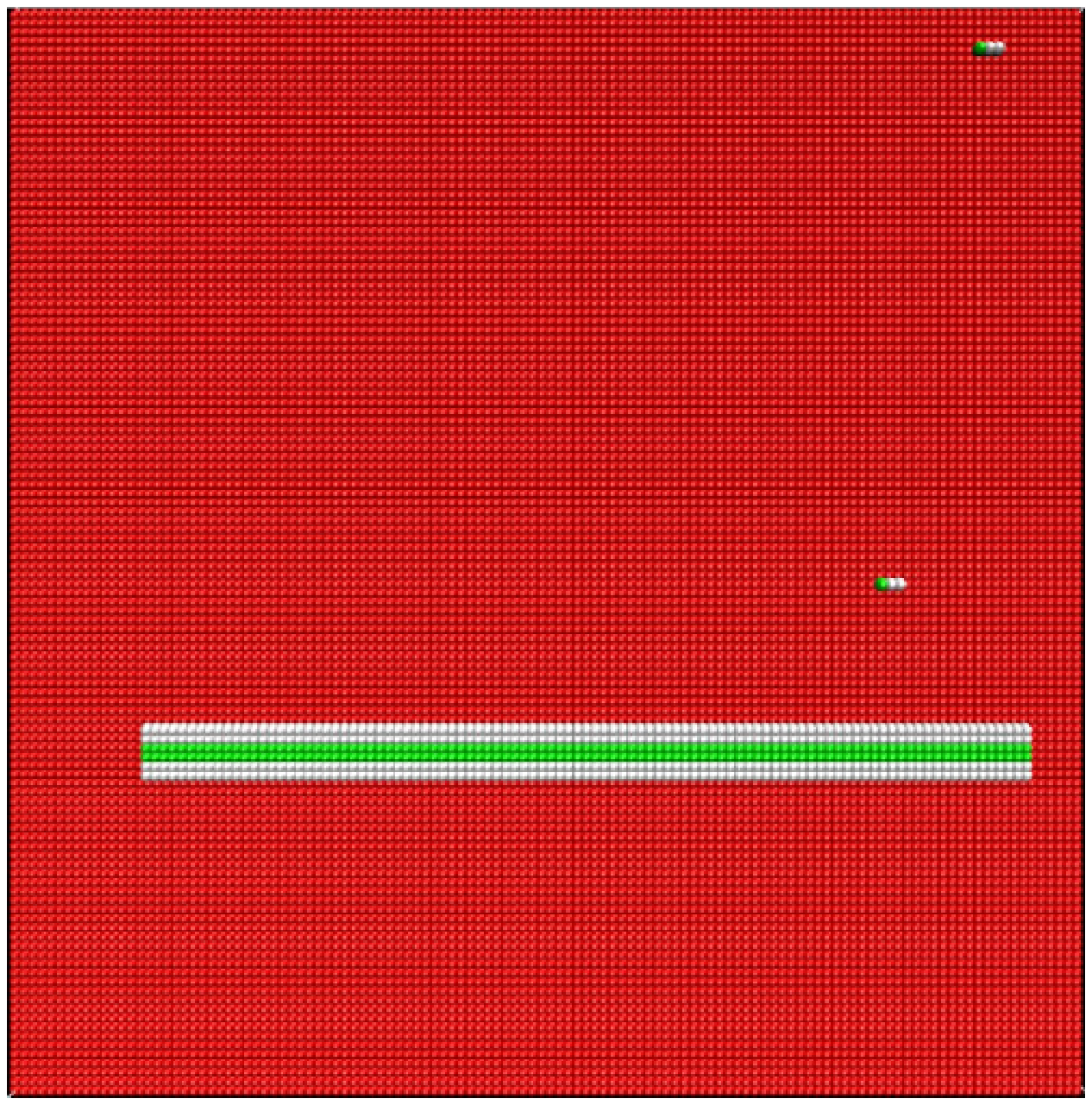} 
    \end{center}

    \begin{center}
      (d)
    \end{center}

     \begin{center}

         \includegraphics[width=0.45\textwidth]{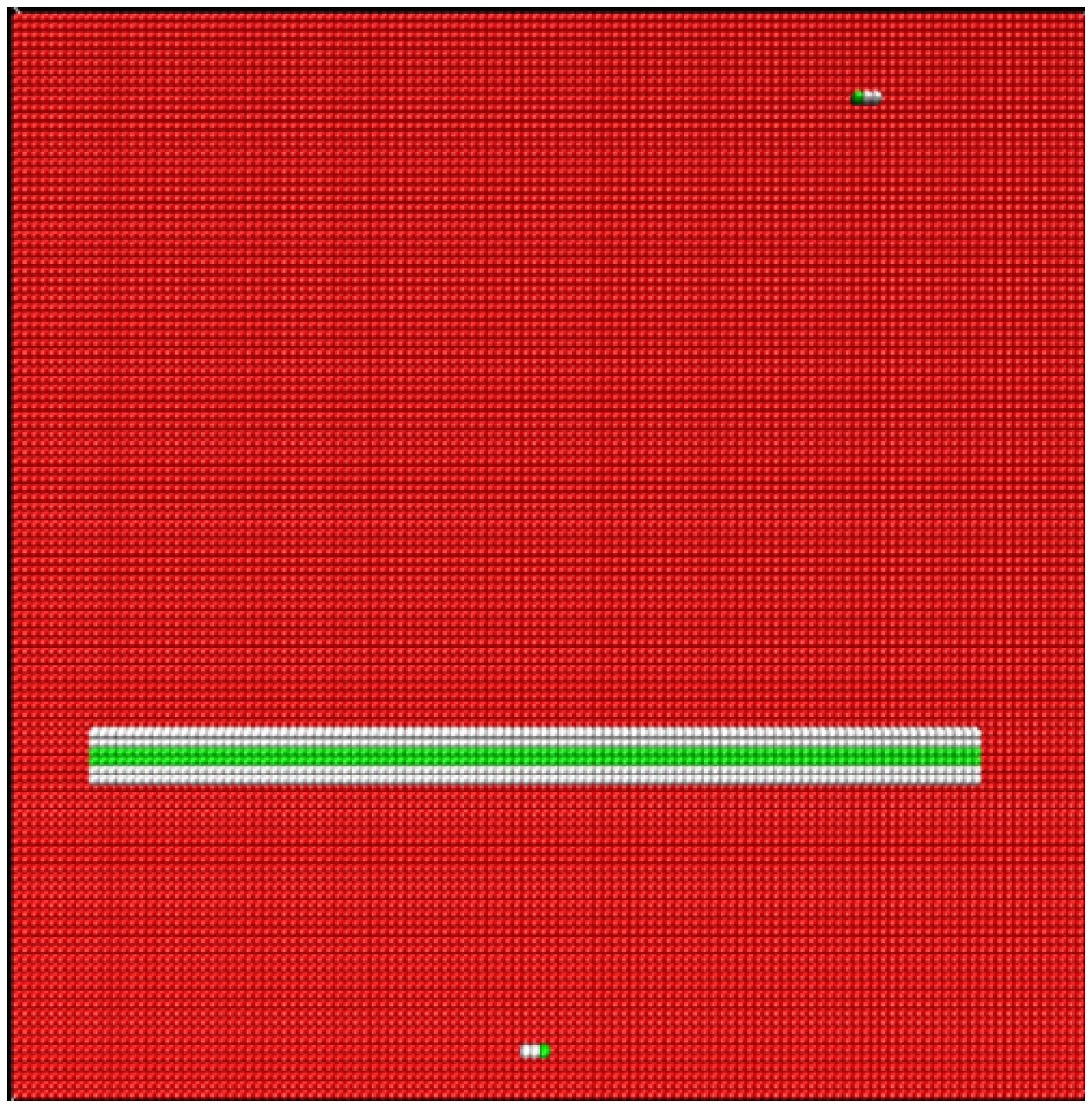}
    \end{center}

    \begin{center}
      (e)
    \end{center}
     \begin{center}

       \includegraphics[width=0.45\textwidth]{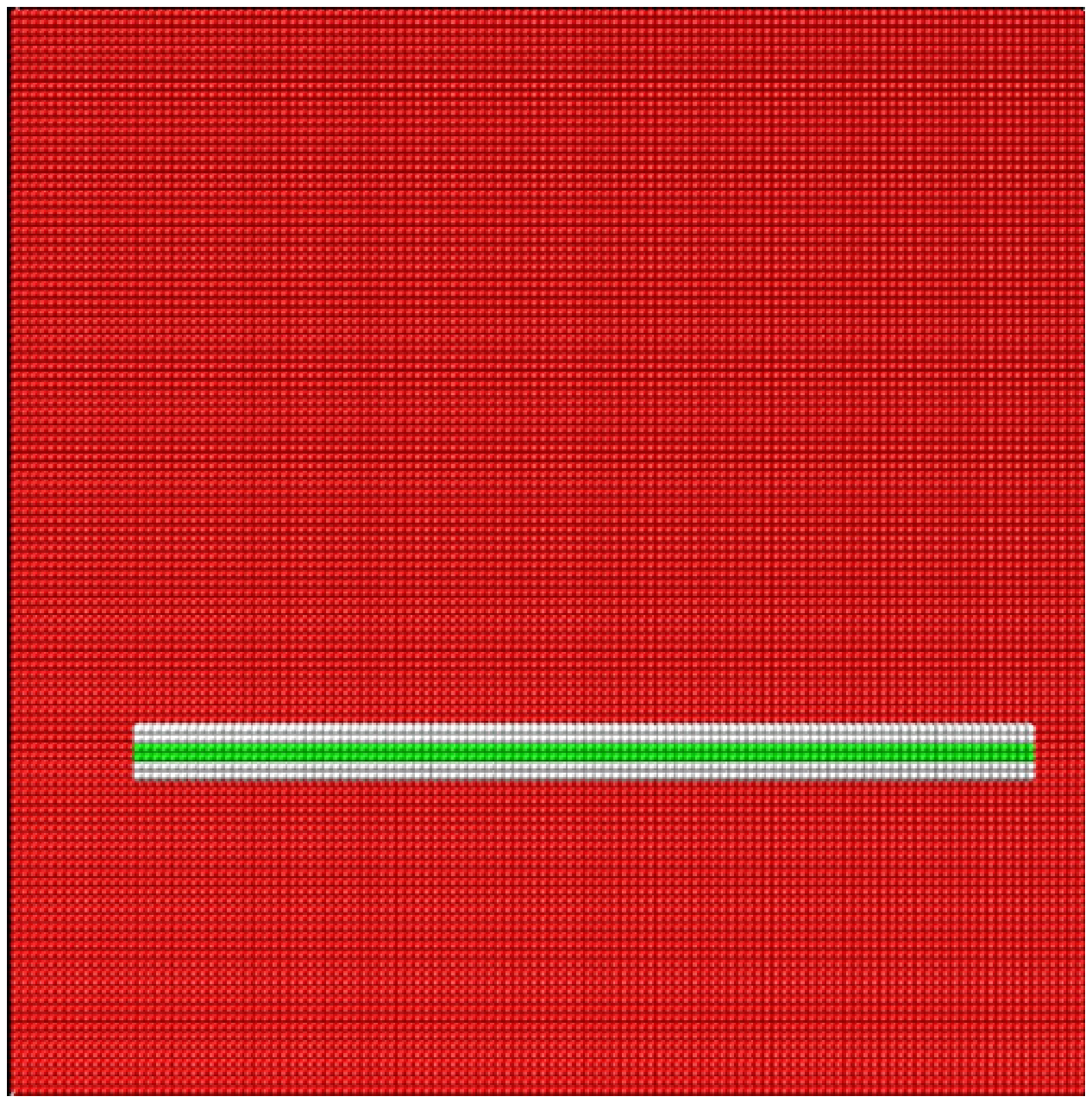}
    \end{center}

    \begin{center}
      (f)
    \end{center}
    \end{minipage}
    \caption{\label {Fig: Mechanism}\small Mechanism to form the final structure. (a), (b) and (c) are the structures at
                                                           $t_{sim}$=0, $t_{sim}$=1$\times$10$^6$ and $t_{sim}$=1$\times$10$^7$
                                                           for $F=$10$^{-2}$.  (d), (e) and (f) are the structures at
                                                           $t_{sim}$=0, $t_{sim}$=1$\times$10$^6$ and $t_{sim}$=1$\times$10$^7$
                                                           for $F=$10$^{-4}$.}
\end{figure}

\begin{figure} [ht]
  \begin{center}
    \includegraphics[width=0.45\textwidth]{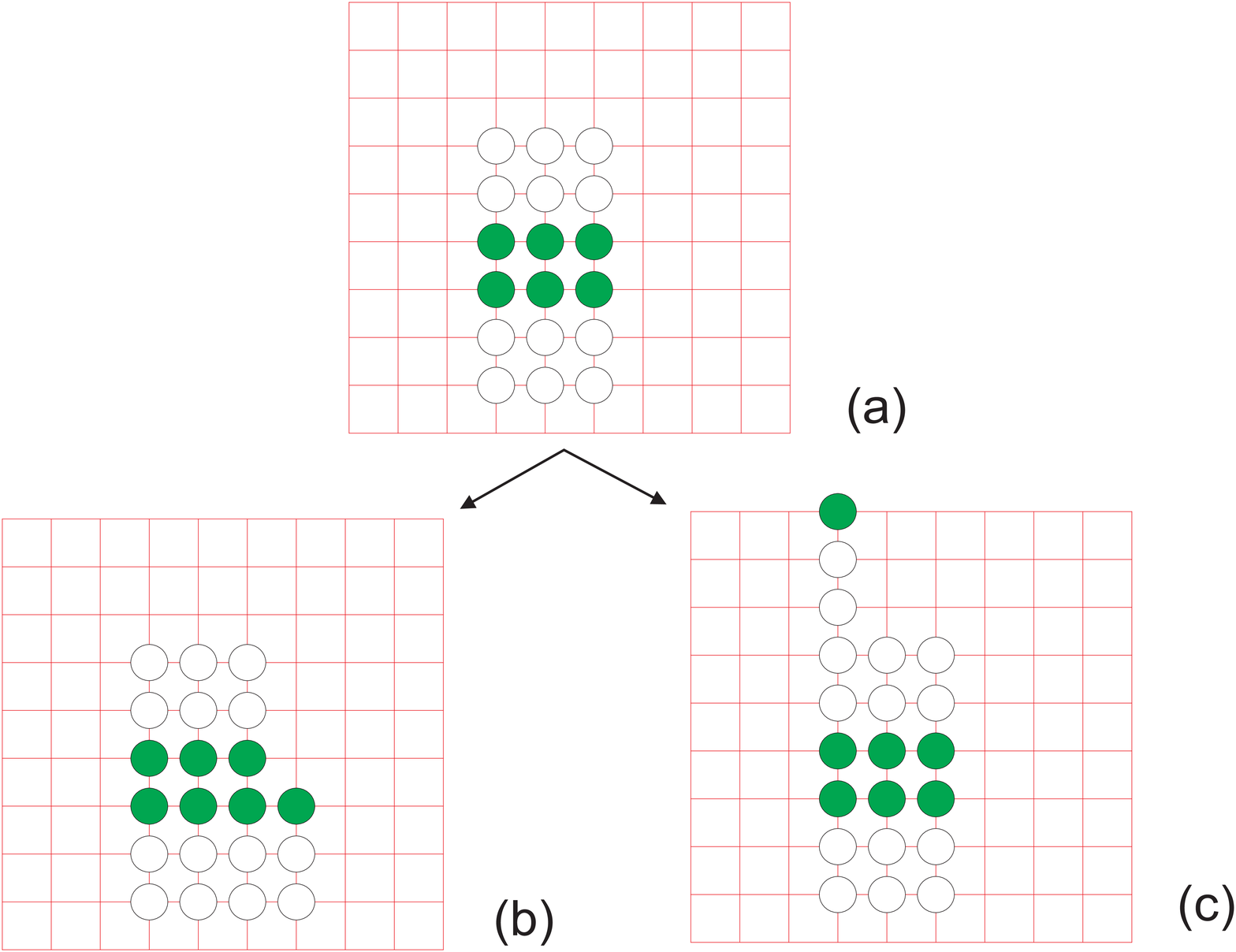}
  \end{center}

\caption{\label {Fig: Trimer}\small (a) The lowest-energy
configuration for 6 trimers. Addition of one trimer leads to
                                    (b) two new TT interactions and one new HH interaction
                                    or (c) one new TT interaction.}

\end{figure}

Here we particularly concentrate on the effect of non-monotonous
flux dependence for a specific temperature. For the temperature of
$T=0.19$ this effect is explicitly highlighted in Fig. \ref{Fig:
E_F} for global as well as local movement. Evidently, one obtains
a significantly non-monotonous dependency of energy on flux. For
$F \le 10^{-4}$ some additional mechanism seems to prevent the
system to reach the same low-energy configurations as compared to
somewhat higher values of the flux. The effect is present for both
the global and the local movement. As expected, via global moves
it is possible to explore the phase space more efficiently, which
explains the differences in energy values for the high flux
regime. Of course, in the limit of extremely small values of the flux any finite system has to reach its equilibrium structure. 
This approach to equilibrium starts to become visible for $F<10^{-5}$; see Fig. \ref{Fig:
E_F}.

Additional insight into this behavior can be gained by analysing
the time-dependence of the energy in the final part of the
simulation; see Fig. \ref {Fig: E_tsim_F}. In the discussion we
concentrate on the comparison of $F=10^{-2}$ and $F=10^{-4}$. For
small values of $t_{sim}$ the configuration, generated with the
high flux of $F=10^{-2}$, is by far more disordered and displays a
much higher energy. However, with increasing simulation time a
typical high-flux configuration manages to approach structures
with energies close to -2.2 whereas the low-flux configuration
seems to be stuck in some region of configuration space because
the energy hardly decreases below -2.13 in the analysed time regime of $t_{sim}$.

To elucidate the reason of this surprising observation we
explicitly monitor the time evolution on a
microscopic scale for both flux values as representative examples, respectively; see
Fig. \ref{Fig: Mechanism}. For the smaller flux one ends up with
the O1-type structure whereas in the other case the final
configuration corresponds to the O2-type structure. How to
understand this behavior? We assume that 6 trimers are present.
The lowest-energy configuration for 6 trimers is shown in Fig.
\ref{Fig: Trimer}(a). As already mentioned above, for a small
number of trimers the O1-configuration has a much lower energy
than the O2-configuration. After the next deposition process the
new trimer will easily find its way to this cluster via diffusion
(local move) or via single hops (global moves). There are,
however, several docking options. Two examples are shown in Fig.
\ref{Fig: Trimer}(b) and (c). Energetically favorable is the
option (b) where the new trimer is fully bound to an already
present trimer. In practice it is likely the new trimer
explores several of the available docking options. If sufficient time is given, 
it is most likely to be found in
configuration (b) but in particular for short times after its
arrival, configuration (c) can also be realized. Here flux enters
into the discussion. With the advent of the next deposited
trimer the present configuration may be stabilized and it will
become more difficult to change the configuration. Thus,
stabilization of configuration (c) is more likely for the high
flux as compared to the low flux. This immediately leads to more
O2-configurations for a high than for a low flux. Of course, several defects may develop 
for a high value of the flux as shown in Fig. \ref{Fig: Mechanism} (a).

\begin{figure} 
  \centering
  \includegraphics[width=0.45\textwidth]{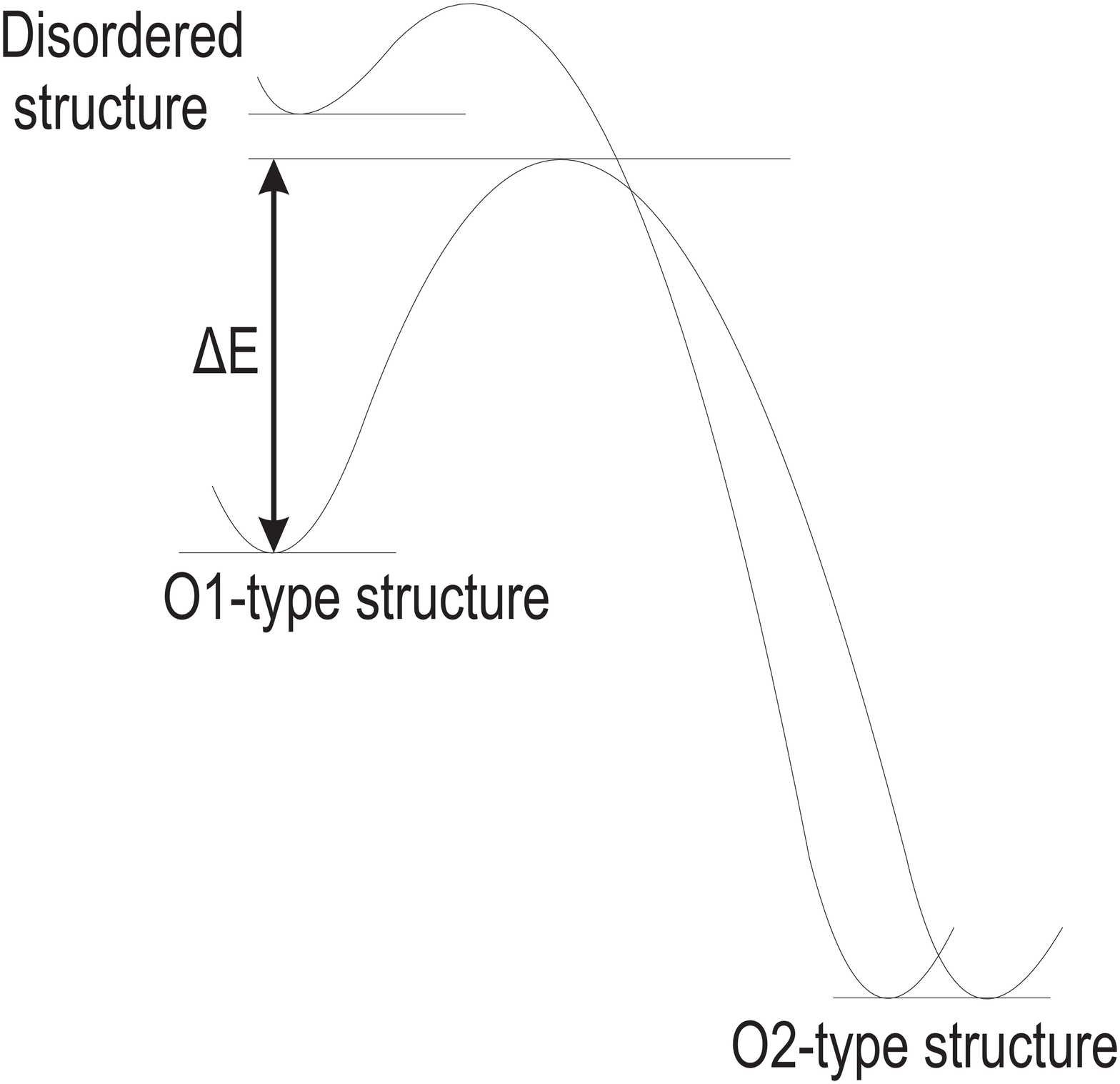}
  \caption{\label{Fig: Activation}\small Schematic presentation of the activation energy for
  structural rearrangement from Figure \ref{Fig: Mechanism} (a) (disordered structure) to
  Figure \ref{Fig: Mechanism} (c) and from Figure \ref{Fig: Mechanism} (d) (O1-type structure) to
  Figure \ref{Fig: Mechanism} (c).}
\end{figure}

The complete qualitative argument for the observed anomalies is
schematically summarized in Fig.\ref{Fig: Activation}. We start with
the low-flux case. As discussed before the system may likely end
up in an O1-structure because this structure is favorable for a
smaller number of trimers. Unfortunately, with an increasing
number of trimers this state  becomes metastable and the
O2-structure becomes the global energy minimum. This barrier is
indicated in Fig. \ref{Fig: Activation}. Starting from a perfectly
ordered O1-structure this transition would involve the successive
desorption of 16 trimers (when the 9$^{th}$ pair of trimers is desorbed, the O2-configuration
becomes energetically favorable) from the O1-structure until the system
realizes that the O2-structure is energetically favorable. As a
consequence neither during the growth period nor during the
evolution period $t_{sim}$ the system can surmount the resulting free
energy barrier to reach the O2-structure. The situation is
different for the high-flux case. During the deposition process it
is likely that beyond the O1-structure additional stripes are
generated, leading to the O2- or even higher order structures.
Additionally, several defects may be present. During the
evolution period $t_{sim}$ the different subclusters typically form one
big cluster. Furthermore, since it is unlikely that the system is
trapped in the O1-configuration it is possible via some minor
activated processes to generate an O2-type structure. In any
event, the resulting free energy barrier is sufficiently small and
can be surmounted during our chosen simulation time $t_{sim}$.

\begin {table}
\begin{center}
  \renewcommand{\arraystretch}{1.2}
    \begin{tabular}{|c|c|c|c|c|c|}
      \hline
          $x$          & 0			&4       		&8    		&12		&16			\\ \hline
      $N_{O1}/N_{O2}$  &1.1		  	&0.3    		&0.2   		&0.1		&0.1			\\ \hline
      $u_{O2}-u_{O1}$  &0.0          		&0.1   			&0.4   		&0.6		&0.6			\\ \hline
      $S_{O1}-S_{O2}$  &0.1$\pm$0.1           	&-1.5$\pm$0.1   	&-3.6$\pm$0.1  	&-4.5$\pm$0.1	&-4.8$\pm$0.1		\\ \hline

    \end{tabular}
  \end{center}
 \caption{\label{Tab: Entropy}\small Thermodynamic properties of O1- and O2-configurations at temperature $T=0.23$ for 36 chains using $L=25$. $x$: Maximum 
number of defects for the definition of O1- and O2-configuration. 
$N_{O1}/N_{O2}$: Probability ratio of forming an O1- as compared to an O2-structure. $u_{O2}-u_{O1}$ : Energy difference between typical O1- and O2-structures.
 $S_{O1}-S_{O2}$: Entropy difference.}
\end{table}

So far we have analysed the energetic aspects of structure formation. Similarly, one may wonder whether entropic aspects favor either the O1- or the O2-configuration. 
Naturally, the perfect configurations have the same entropy. At finite temperatures both types of configurations may have some defects. The number of defects with still 
reasonable energies may be different for both types of configurations. Formally, this can be expressed in terms of an equilibrium entropy. Therefore we have performed 
long simulations at $T=0.23$ (using $L=0.25$) where we determined the probability that a O1- or O2-configuration with at most x ($x \in \{0,4,8,12,16\})$ defects occurs. Using the standard relation $Z = \exp(U-TS)$ 
we have obtained the relative entropies ($S_{O1}-S_{O2}$) from knowledge of the relative populations ($Z_{O1}/Z_{O2}$) and the average energies ($u_{O2}-u_{O1}$) . 
For this specific simulation we have used $N=36$ because for this system size the perfect O1- and O2-structure have nearly identical energies. The results are listed in Tab.\ref{Tab: Entropy}. 
Indeed one can see that entropically the O2-structure is favored. The entropy difference may thus help  to transfer an O1-structure to an O2-structure because of larger attraction basin of the O2-structure.

\begin{figure}
  \centering
  \includegraphics[width=0.45\textwidth]{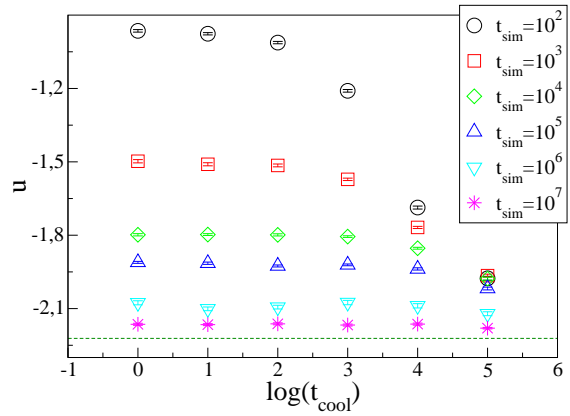}
  \caption{\label {Fig: E_tcool} \small Change of $u$ with cooling rate for different choices of $t_{sim}$ at $T$=0.19.
                                        The dotted horizontal line indicates the minimum $u$ of the system.}
\end{figure}

\begin{figure}
  \centering
  \includegraphics[width=0.45\textwidth]{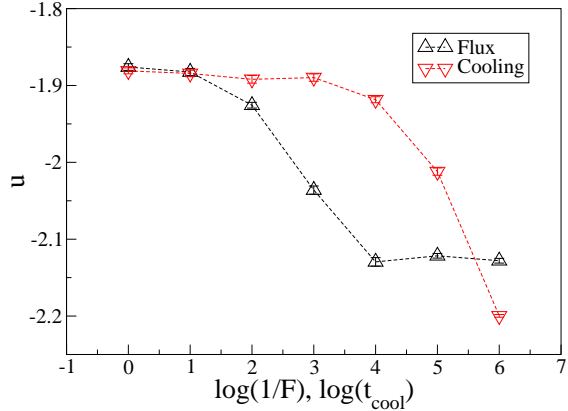}
\begin{center}
 (a)
\end{center}

  \centering
    \includegraphics[width=0.45\textwidth]{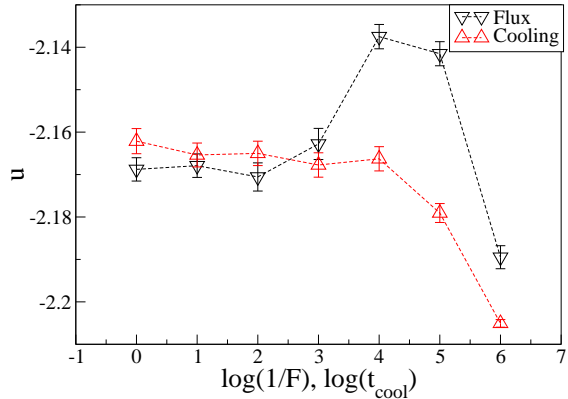}
\begin{center}
 (b)
\end{center}
  \caption{\label {Fig: CoolingvsFlux} \small Comparison of the resulting energy for the flux simulations
                                         and the cooling simulations at (a) $T=0.05$ and (b) $T=0.19$. The dashed line is a guide to the eyes.}
\end{figure}

\subsection{Cooling simulations}
In the next step, we switch from the  deposition to the cooling mechanism
to reach the final values of $N$ and $T$. Choosing a final
temperature of $T=0.19$ we just observe the standard behavior; see
Fig. \ref{Fig: E_tcool}. First, shortly after the cooling period
is finished the configurations generated with the fastest cooling
protocol have the highest energy. Second, after the cooling when the evolution
period $t_{sim}$ is finished, the energy still depends in
a monotonous way on the initial cooling rate. The difference can
be easily rationalized. Since from the very beginning all $N=200$
chains are present just for entropic reasons the system will not
be confined to end up in the O1-structure. Naturally, for given
cooling time $t_{cool}$ aging effects can be only observed if
$t_{sim}$ starts to exceed the initial cooling time $t_{cool}$.

Finally, it may be interesting to compare the degree of
equilibration for the flux and the cooling setup. For this purpose
we monitor the energy at T=0.19 and T=0.05 for both scenarios. By
matching the respective times of the flux and the cooling period,
respectively, we can directly compare the efficiency of both
methods to reach structures with lower energies in the
low-temperature regime; see Fig. \ref{Fig: CoolingvsFlux}. It turns out that for the main
part of the parameter regime the flux simulations are  more
efficient.  This effect is particularly pronounced at very low
temperatures. Since the cooling simulation starts with all $N=200$
chains very complex disordered structures may be generated which
cannot be dissolved during the simulation. In contrast, due to the
gradual increase of the number of molecules in the initial part of
the flux simulation this effect seems to be less pronounced. The
results are different in the regime where the O1- vs. O2-problem,
discussed in this work, becomes relevant. As one can see for
$F=10^{-4}$ and $F=10^{-5}$ at $T=0.19$
cooling simulations are more efficient. This effect can easily  be
explained. In the flux simulations one generates with a high
probability  O1-configurations which, finally gives rise to
somewhat higher energies as compared to the simulations with
higher flux. Naturally, this effect does not occur for the cooling
simulations because the system starts with a very disordered
structure at the starting temperature $T_h$.

\section{\label{sec:summary}Summary and conclusion}
With the help of Monte Carlo simulation we have analyzed the
effect of flux and temperature on the behavior of a monolayer film
formed by rigid head-tail trimers on the square lattice. Our
chosen evolution time $t_{sim}$ was long enough so that for $T
\ge 0.23$ the final configuration corresponds to an equilibration
configuration, i.e. its energy does not depend on the
initialization period.  The most interesting temperature regime is
slightly below the onset of equilibration. Here we find an
increase of the final energy with decreasing flux. On a
qualitative level this effect can be understood from the
observation that for very low flux the system ends up in a O1-type
configuration which for a small number of chains forms the global
energy minimum. However, with increasing number of adsorbed chains
this global minimum starts to become a local minimum. Then the
system is no longer able to escape to the newly developed global
minimum, corresponding to the O2-type configuration. In contrast,
when replacing the deposition period by a cooling period the
expected correlation between energy and cooling rate is observed.

We would like to note that the same scenario is also obtained for larger chains because this general mechanism works as well. 
For example we have observed that for a hexamer (one H and five T groups) the energy also displays a maximum as a function of 
temperature using $F=10^{-5}$ and $t_{sim}=5 \times 10^7$. It is observed at $T=0.25$. In general, the critical temperature range as well as the critical cluster
size naturally depends on the potential parameters and the chain length. For example critical cluster size which is $N=36$ for our standard trimer increases to $N=60$ for this hexamer.

When the system size is doubled analogous effects are also observed. Indeed, since the mechanism of this anomalous flux dependence is very general, it is not surprising that
this phenomenon is robust against variation of system parameters. Of course, going to even much larger values of $L$, it would be required to increase the number of chains accordingly. Then one would observe 
several smaller clusters. Their local properties, however, just follow the general mechanisms described in this work.

This work shows that the observation of Ediger and coworkers about
the monotonous enthalpy-flux dependence does not hold in general
\cite{Ediger}. Rather for molecules with intrinsic anisotropies a
more complex behavior can occur. However, in agreement with these
experiments  we generally observe that the system generation via
deposition can be very efficient to generate low-energy structures
if compared with cooling from a high-temperature equilibrium structure.

By the same experimental setup from Ref.\cite{Ediger}, in principle it may be
possible to observe these effects also experimentally
for chain molecules, containing a head- and a short tail-region. The only condition is the separation of energy scales of the two types of binding as shown in Fig. \ref{Fig: Trimer}.
The present simulations indicate that the temperature regime for
this effect may be sufficiently large. For our specific choice of
parameters  it is approx. between the temperatures 0.15 and 0.2.
In further work it may be helpful to study even simpler model
systems where also analytical calculations are possible. Then it
may be checked in more detail for which interaction parameters an
optimum visibility of this anomalous flux-dependence is possible.
This may be helpful for the question for which real-world systems
this effect may also be seen experimentally.

\section{Acknowledgement}

We gratefully acknowledge the support by the DFG (SFB 858) and
helpful discussions with Mark Ediger and Lifeng Chi as well as her
group.


\begin{thebibliography}{10}%
\makeatletter
\providecommand \@ifxundefined [1]{%
 \ifx #1\undefined \expandafter \@firstoftwo
 \else \expandafter \@secondoftwo
\fi
}%
\providecommand \@ifnum [1]{%
 \ifnum #1\expandafter \@firstoftwo
 \else \expandafter \@secondoftwo
\fi
}%
\providecommand \enquote [1]{``#1''}%
\providecommand \bibnamefont  [1]{#1}%
\providecommand \bibfnamefont [1]{#1}%
\providecommand \citenamefont [1]{#1}%
\providecommand\href[0]{\@sanitize\@href}%
\providecommand\@href[1]{\endgroup\@@startlink{#1}\endgroup\@@href}%
\providecommand\@@href[1]{#1\@@endlink}%
\providecommand \@sanitize [0]{\begingroup\catcode`\&12\catcode`\#12\relax}%
\@ifxundefined \pdfoutput {\@firstoftwo}{%
 \@ifnum{\z@=\pdfoutput}{\@firstoftwo}{\@secondoftwo}%
}{%
 \providecommand\@@startlink[1]{\leavevmode}%
 \providecommand\@@endlink[0]{}%
}{%
 \providecommand\@@startlink[1]{%
  \leavevmode
  \pdfstartlink
   attr{/Border[0 0 1 ]/H/I/C[0 1 1]}%
   user{/Subtype/Link/A<</Type/Action/S/URI/URI(#1)>>}%
  \relax
 }%
 \providecommand\@@endlink[0]{\pdfendlink}%
}%
\providecommand \url  [0]{\begingroup\@sanitize \@url }%
\providecommand \@url [1]{\endgroup\@href {#1}{\urlprefix}}%
\providecommand \urlprefix [0]{URL }%
\providecommand \Eprint[0]{\href }%
\@ifxundefined \urlstyle {%
  \providecommand \doi [1]{doi:\discretionary{}{}{}#1}%
}{%
  \providecommand \doi [0]{doi:\discretionary{}{}{}\begingroup
  \urlstyle{rm}\Url }%
}%
\providecommand \doibase [0]{http://dx.doi.org/}%
\providecommand \Doi[1]{\href{\doibase#1}}%
\providecommand \selectlanguage [0]{\@gobble}%
\providecommand \bibinfo [0]{\@secondoftwo}%
\providecommand \bibfield [0]{\@secondoftwo}%
\providecommand \translation [1]{[#1]}%
\providecommand \BibitemOpen[0]{}%
\providecommand \bibitemStop [0]{}%
\providecommand \bibitemNoStop [0]{.\EOS\space}%
\providecommand \EOS [0]{\spacefactor3000\relax}%
\providecommand \BibitemShut [1]{\csname bibitem#1\endcsname}%
\bibitem{Krug}%
  \BibitemOpen
  \bibfield{author}{%
  \bibinfo {author} {\bibfnamefont{T.}~\bibnamefont{Michely}}\ and\ \bibinfo
  {author} {\bibfnamefont{J.}~\bibnamefont{Krug}},\ }%
  \emph{\bibinfo {title} {Islands, Mounds and Atoms}}\ (\bibinfo {publisher}
  {Springer-Verlag},\ \bibinfo {address} {Berlin},\ \bibinfo {year}
  {2004})\BibitemShut{NoStop}%
\bibitem{Wang}%
  \BibitemOpen
  \bibfield{author}{%
  \bibinfo {author} {\bibfnamefont{W.~C.}\ \bibnamefont{Wang}}, \bibinfo
  {author} {\bibfnamefont{D.~Y.}\ \bibnamefont{Zhong}}, \bibinfo {author}
  {\bibfnamefont{J.}~\bibnamefont{Zhu}}, \bibinfo {author}
  {\bibfnamefont{F.}~\bibnamefont{Kalischewski}}, \bibinfo {author}
  {\bibfnamefont{R.~F.}\ \bibnamefont{Dou}}, \bibinfo {author}
  {\bibfnamefont{K.}~\bibnamefont{Wedeking}}, \bibinfo {author}
  {\bibfnamefont{Y.}~\bibnamefont{Wang}}, \bibinfo {author}
  {\bibfnamefont{A.}~\bibnamefont{Heuer}}, \bibinfo {author}
  {\bibfnamefont{H.}~\bibnamefont{Fuchs}}, \bibinfo {author}
  {\bibfnamefont{G.}~\bibnamefont{Erker}},\ and\ \bibinfo {author}
  {\bibfnamefont{L.~F.}\ \bibnamefont{Chi}},\ }%
  \bibfield{journal}{%
  \bibinfo {journal} {Phys. Rev. Lett.}\ }%
  \textbf{\bibinfo {volume} {98}},\ \bibinfo {pages} {225504} (\bibinfo {year}
  {2007})\BibitemShut{NoStop}%
\bibitem{Venables}%
  \BibitemOpen
  \bibfield{author}{%
  \bibinfo {author} {\bibfnamefont{J.~A.}\ \bibnamefont{Venables}}\ and\
  \bibinfo {author} {\bibfnamefont{J.~H.}\ \bibnamefont{Harding}},\ }%
  \bibfield{journal}{%
  \bibinfo {journal} {J. Cryst. Growth}\ }%
  \textbf{\bibinfo {volume} {211}},\ \bibinfo {pages} {27 } (\bibinfo {year}
  {2000})\BibitemShut{NoStop}%
\bibitem{Haas}%
  \BibitemOpen
  \bibfield{author}{%
  \bibinfo {author} {\bibfnamefont{G.}~\bibnamefont{Haas}}, \bibinfo {author}
  {\bibfnamefont{A.}~\bibnamefont{Menck}}, \bibinfo {author}
  {\bibfnamefont{H.}~\bibnamefont{Brune}}, \bibinfo {author}
  {\bibfnamefont{J.~V.}\ \bibnamefont{Barth}}, \bibinfo {author}
  {\bibfnamefont{J.~A.}\ \bibnamefont{Venables}},\ and\ \bibinfo {author}
  {\bibfnamefont{K.}~\bibnamefont{Kern}},\ }%
  \bibfield{journal}{%
  \bibinfo {journal} {Phys. Rev. B}\ }%
  \textbf{\bibinfo {volume} {61}},\ \bibinfo {pages} {11105} (\bibinfo {year}
  {2000})\BibitemShut{NoStop}%
\bibitem{Vardavas}%
  \BibitemOpen
  \bibfield{author}{%
  \bibinfo {author} {\bibfnamefont{R.}~\bibnamefont{Vardavas}}, \bibinfo
  {author} {\bibfnamefont{C.}~\bibnamefont{Ratsch}},\ and\ \bibinfo {author}
  {\bibfnamefont{R.~E.}\ \bibnamefont{Caflisch}},\ }%
  \bibfield{journal}{%
  \bibinfo {journal} {Surf. Sci.}\ }%
  \textbf{\bibinfo {volume} {569}},\ \bibinfo {pages} {185 } (\bibinfo {year}
  {2004})\BibitemShut{NoStop}%
\bibitem{Nurminen}%
  \BibitemOpen
  \bibfield{author}{%
  \bibinfo {author} {\bibfnamefont{L.}~\bibnamefont{Nurminen}}, \bibinfo
  {author} {\bibfnamefont{A.}~\bibnamefont{Kuronen}},\ and\ \bibinfo {author}
  {\bibfnamefont{K.}~\bibnamefont{Kaski}},\ }%
  \bibfield{journal}{%
  \bibinfo {journal} {Phys. Rev. B}\ }%
  \textbf{\bibinfo {volume} {63}},\ \bibinfo {pages} {035407} (\bibinfo {year}
  {2000})\BibitemShut{NoStop}%
\bibitem{Sabiryanov}%
  \BibitemOpen
  \bibfield{author}{%
  \bibinfo {author} {\bibfnamefont{R.~F.}\ \bibnamefont{Sabiryanov}}, \bibinfo
  {author} {\bibfnamefont{M.~I.}\ \bibnamefont{Larsson}}, \bibinfo {author}
  {\bibfnamefont{K.~J.}\ \bibnamefont{Cho}}, \bibinfo {author}
  {\bibfnamefont{W.~D.}\ \bibnamefont{Nix}},\ and\ \bibinfo {author}
  {\bibfnamefont{B.~M.}\ \bibnamefont{Clemens}},\ }%
  \bibfield{journal}{%
  \bibinfo {journal} {Phys. Rev. B}\ }%
  \textbf{\bibinfo {volume} {67}},\ \bibinfo {pages} {125412} (\bibinfo {year}
  {2003})\BibitemShut{NoStop}%
\bibitem{zhong}%
  \BibitemOpen
  \bibfield{author}{%
  \bibinfo {author} {\bibfnamefont{Z.}~\bibnamefont{Zhong}}\ and\ \bibinfo
  {author} {\bibfnamefont{G.}~\bibnamefont{Bauer}},\ }%
  \bibfield{journal}{%
  \bibinfo {journal} {Appl. Phys. Lett.}\ }%
  \textbf{\bibinfo {volume} {84}},\ \bibinfo {pages} {1922} (\bibinfo {year}
  {2004})\BibitemShut{NoStop}%
\bibitem{Briseno}%
  \BibitemOpen
  \bibfield{author}{%
  \bibinfo {author} {\bibfnamefont{A.~L.}\ \bibnamefont{Briseno}}, \bibinfo
  {author} {\bibfnamefont{J.}~\bibnamefont{Aizenberg}}, \bibinfo {author}
  {\bibfnamefont{Y.-J.}\ \bibnamefont{Han}}, \bibinfo {author}
  {\bibfnamefont{R.~A.}\ \bibnamefont{Penkala}}, \bibinfo {author}
  {\bibfnamefont{H.}~\bibnamefont{Moon}}, \bibinfo {author}
  {\bibfnamefont{A.~J.}\ \bibnamefont{Lovinger}}, \bibinfo {author}
  {\bibfnamefont{C.}~\bibnamefont{Kloc}},\ and\ \bibinfo {author}
  {\bibfnamefont{Z.}~\bibnamefont{Bao}},\ }%
  \bibfield{journal}{%
  \bibinfo {journal} {J. Am. Chem. Soc.}\ }%
  \textbf{\bibinfo {volume} {127}},\ \bibinfo {pages} {12164} (\bibinfo {year}
  {2005})\BibitemShut{NoStop}%
\bibitem{Fuchs}%
  \BibitemOpen
  \bibfield{author}{%
  \bibinfo {author} {\bibfnamefont{B.}~\bibnamefont{Dong}}, \bibinfo {author}
  {\bibfnamefont{D.~Y.}\ \bibnamefont{Zhong}}, \bibinfo {author}
  {\bibfnamefont{L.~F.}\ \bibnamefont{Chi}},\ and\ \bibinfo {author}
  {\bibfnamefont{H.}~\bibnamefont{Fuchs}},\ }%
  \bibfield{journal}{%
  \bibinfo {journal} {Adv. Mater.}\ }%
  \textbf{\bibinfo {volume} {17}},\ \bibinfo {pages} {2736} (\bibinfo {year}
  {2005})\BibitemShut{NoStop}%
\bibitem{Kalischewski_1}%
  \BibitemOpen
  \bibfield{author}{%
  \bibinfo {author} {\bibfnamefont{F.}~\bibnamefont{Kalischewski}}, \bibinfo
  {author} {\bibfnamefont{J.}~\bibnamefont{Zhu}},\ and\ \bibinfo {author}
  {\bibfnamefont{A.}~\bibnamefont{Heuer}},\ }%
  \bibfield{journal}{%
  \bibinfo {journal} {Phys. Rev. B}\ }%
  \textbf{\bibinfo {volume} {78}},\ \bibinfo {pages} {155401} (\bibinfo {year}
  {2008})\BibitemShut{NoStop}%
\bibitem{Kalischewski_2}%
  \BibitemOpen
  \bibfield{author}{%
  \bibinfo {author} {\bibfnamefont{F.}~\bibnamefont{Kalischewski}}\ and\
  \bibinfo {author} {\bibfnamefont{A.}~\bibnamefont{Heuer}},\ }%
  \bibfield{journal}{%
  \bibinfo {journal} {Phys. Rev. B}\ }%
  \textbf{\bibinfo {volume} {80}},\ \bibinfo {pages} {155421} (\bibinfo {year}
  {2009})\BibitemShut{NoStop}%
\bibitem{Lied}%
  \BibitemOpen
  \bibfield{author}{%
  \bibinfo {author} {\bibfnamefont{F.}~\bibnamefont{Lied}}, \bibinfo {author}
  {\bibfnamefont{T.}~\bibnamefont{Mues}}, \bibinfo {author}
  {\bibfnamefont{W.}~\bibnamefont{Wang}}, \bibinfo {author}
  {\bibfnamefont{L.}~\bibnamefont{Chi}},\ and\ \bibinfo {author}
  {\bibfnamefont{A.}~\bibnamefont{Heuer}},\ }%
  \bibfield{journal}{%
  \bibinfo {journal} {J. Chem. Phys.}\ }%
  \textbf{\bibinfo {volume} {136}},\ \bibinfo {pages} {024704} (\bibinfo {year}
  {2012})\BibitemShut{NoStop}%
\bibitem{tsumura}%
  \BibitemOpen
  \bibfield{author}{%
  \bibinfo {author} {\bibfnamefont{A.}~\bibnamefont{Tsumura}}, \bibinfo
  {author} {\bibfnamefont{H.}~\bibnamefont{Koezuka}},\ and\ \bibinfo {author}
  {\bibfnamefont{T.}~\bibnamefont{Ando}},\ }%
  \bibfield{journal}{%
  \bibinfo {journal} {Appl. Phys. Lett.}\ }%
  \textbf{\bibinfo {volume} {49}},\ \bibinfo {pages} {1210} (\bibinfo {year}
  {1986})\BibitemShut{NoStop}%
\bibitem{Burroughes_1}%
  \BibitemOpen
  \bibfield{author}{%
  \bibinfo {author} {\bibfnamefont{J.~H.}\ \bibnamefont{Burroughes}}, \bibinfo
  {author} {\bibfnamefont{C.~A.}\ \bibnamefont{Jones}},\ and\ \bibinfo {author}
  {\bibfnamefont{R.~H.}\ \bibnamefont{Friend}},\ }%
  \bibfield{journal}{%
  \bibinfo {journal} {Nature}\ }%
  \textbf{\bibinfo {volume} {335}},\ \bibinfo {pages} {137} (\bibinfo {year}
  {1988})\BibitemShut{NoStop}%
\bibitem{assadi}%
  \BibitemOpen
  \bibfield{author}{%
  \bibinfo {author} {\bibfnamefont{A.}~\bibnamefont{Assadi}}, \bibinfo {author}
  {\bibfnamefont{C.}~\bibnamefont{Svensson}}, \bibinfo {author}
  {\bibfnamefont{M.}~\bibnamefont{Willander}},\ and\ \bibinfo {author}
  {\bibfnamefont{O.}~\bibnamefont{Ingan\"{a}s}},\ }%
  \bibfield{journal}{%
  \bibinfo {journal} {Appl. Phys. Lett.}\ }%
  \textbf{\bibinfo {volume} {53}},\ \bibinfo {pages} {195} (\bibinfo {year}
  {1988})\BibitemShut{NoStop}%
\bibitem{paloheimo}%
  \BibitemOpen
  \bibfield{author}{%
  \bibinfo {author} {\bibfnamefont{J.}~\bibnamefont{Paloheimo}}, \bibinfo
  {author} {\bibfnamefont{P.}~\bibnamefont{Kuivalainen}}, \bibinfo {author}
  {\bibfnamefont{H.}~\bibnamefont{Stubb}}, \bibinfo {author}
  {\bibfnamefont{E.}~\bibnamefont{Vuorimaa}},\ and\ \bibinfo {author}
  {\bibfnamefont{P.}~\bibnamefont{Yli-Lahti}},\ }%
  \bibfield{journal}{%
  \bibinfo {journal} {Appl. Phys. Lett.}\ }%
  \textbf{\bibinfo {volume} {56}},\ \bibinfo {pages} {1157} (\bibinfo {year}
  {1990})\BibitemShut{NoStop}%
\bibitem{Horowitz}%
  \BibitemOpen
  \bibfield{author}{%
  \bibinfo {author} {\bibfnamefont{G.}~\bibnamefont{Horowitz}}, \bibinfo
  {author} {\bibfnamefont{D.}~\bibnamefont{Fichou}}, \bibinfo {author}
  {\bibfnamefont{X.}~\bibnamefont{Peng}}, \bibinfo {author}
  {\bibfnamefont{Z.}~\bibnamefont{Xu}},\ and\ \bibinfo {author}
  {\bibfnamefont{F.}~\bibnamefont{Garnier}},\ }%
  \bibfield{journal}{%
  \bibinfo {journal} {Solid State Commun.}\ }%
  \textbf{\bibinfo {volume} {72}},\ \bibinfo {pages} {381 } (\bibinfo {year}
  {1989})\BibitemShut{NoStop}%
\bibitem{Burroughes_2}%
  \BibitemOpen
  \bibfield{author}{%
  \bibinfo {author} {\bibfnamefont{J.~H.}\ \bibnamefont{Burroughes}}, \bibinfo
  {author} {\bibfnamefont{D.~D.~C.}\ \bibnamefont{Bradley}}, \bibinfo {author}
  {\bibfnamefont{A.~R.}\ \bibnamefont{Brown}}, \bibinfo {author}
  {\bibfnamefont{R.~N.}\ \bibnamefont{Marks}}, \bibinfo {author}
  {\bibfnamefont{K.}~\bibnamefont{Mackay}}, \bibinfo {author}
  {\bibfnamefont{R.~H.}\ \bibnamefont{Friend}}, \bibinfo {author}
  {\bibfnamefont{P.~L.}\ \bibnamefont{Burns}},\ and\ \bibinfo {author}
  {\bibfnamefont{A.~B.}\ \bibnamefont{Holmes}},\ }%
  \bibfield{journal}{%
  \bibinfo {journal} {Nature}\ }%
  \textbf{\bibinfo {volume} {347}},\ \bibinfo {pages} {539} (\bibinfo {year}
  {1990})\BibitemShut{NoStop}%
\bibitem{YutakaOhmori}%
  \BibitemOpen
  \bibfield{author}{%
  \bibinfo {author} {\bibfnamefont{Y.}~\bibnamefont{Ohmori}}, \bibinfo {author}
  {\bibfnamefont{M.}~\bibnamefont{Uchida}}, \bibinfo {author}
  {\bibfnamefont{K.}~\bibnamefont{Muro}},\ and\ \bibinfo {author}
  {\bibfnamefont{K.}~\bibnamefont{Yoshino}},\ }%
  \bibfield{journal}{%
  \bibinfo {journal} {Solid State Commun.}\ }%
  \textbf{\bibinfo {volume} {80}},\ \bibinfo {pages} {605 } (\bibinfo {year}
  {1991})\BibitemShut{NoStop}%
\bibitem{braun}%
  \BibitemOpen
  \bibfield{author}{%
  \bibinfo {author} {\bibfnamefont{D.}~\bibnamefont{Braun}}\ and\ \bibinfo
  {author} {\bibfnamefont{A.~J.}\ \bibnamefont{Heeger}},\ }%
  \bibfield{journal}{%
  \bibinfo {journal} {Appl. Phys. Lett.}\ }%
  \textbf{\bibinfo {volume} {58}},\ \bibinfo {pages} {1982} (\bibinfo {year}
  {1991})\BibitemShut{NoStop}%
\bibitem{Yu}%
  \BibitemOpen
  \bibfield{author}{%
  \bibinfo {author} {\bibfnamefont{G.}~\bibnamefont{Yu}}, \bibinfo {author}
  {\bibfnamefont{J.}~\bibnamefont{Gao}}, \bibinfo {author}
  {\bibfnamefont{J.~C.}\ \bibnamefont{Hummelen}}, \bibinfo {author}
  {\bibfnamefont{F.}~\bibnamefont{Wudl}},\ and\ \bibinfo {author}
  {\bibfnamefont{A.~J.}\ \bibnamefont{Heeger}},\ }%
  \bibfield{journal}{%
  \bibinfo {journal} {Science}\ }%
  \textbf{\bibinfo {volume} {270}},\ \bibinfo {pages} {1789} (\bibinfo {year}
  {1995})\BibitemShut{NoStop}%
\bibitem{Sirringhaus}%
  \BibitemOpen
  \bibfield{author}{%
  \bibinfo {author} {\bibfnamefont{H.}~\bibnamefont{Sirringhaus}}, \bibinfo
  {author} {\bibfnamefont{N.}~\bibnamefont{Tessler}},\ and\ \bibinfo {author}
  {\bibfnamefont{R.~H.}\ \bibnamefont{Friend}},\ }%
  \bibfield{journal}{%
  \bibinfo {journal} {Science}\ }%
  \textbf{\bibinfo {volume} {280}},\ \bibinfo {pages} {1741} (\bibinfo {year}
  {1998})\BibitemShut{NoStop}%
\bibitem{Yang}%
  \BibitemOpen
  \bibfield{author}{%
  \bibinfo {author} {\bibfnamefont{J.-S.}\ \bibnamefont{Yang}}\ and\ \bibinfo
  {author} {\bibfnamefont{T.~M.}\ \bibnamefont{Swager}},\ }%
  \bibfield{journal}{%
  \bibinfo {journal} {J. Am. Chem. Soc.}\ }%
  \textbf{\bibinfo {volume} {120}},\ \bibinfo {pages} {11864} (\bibinfo {year}
  {1998})\BibitemShut{NoStop}%
\bibitem{Hopp}%
  \BibitemOpen
  \bibfield{author}{%
  \bibinfo {author} {\bibfnamefont{S.~F.}\ \bibnamefont{Hopp}}\ and\ \bibinfo
  {author} {\bibfnamefont{A.}~\bibnamefont{Heuer}},\ }%
  \bibfield{journal}{%
  \bibinfo {journal} {J. Chem. Phys.}\ }%
  \textbf{\bibinfo {volume} {136}},\ \bibinfo {pages} {154106} (\bibinfo
  {month} {Apr}\ \bibinfo {year} {2012})\BibitemShut{NoStop}%
\bibitem{bellier-castella}%
  \BibitemOpen
  \bibfield{author}{%
  \bibinfo {author} {\bibfnamefont{L.}~\bibnamefont{Bellier-Castella}},
  \bibinfo {author} {\bibfnamefont{D.}~\bibnamefont{Caprion}},\ and\ \bibinfo
  {author} {\bibfnamefont{J.-P.}\ \bibnamefont{Ryckaert}},\ }%
  \bibfield{journal}{%
  \bibinfo {journal} {J. Chem. Phys.}\ }%
  \textbf{\bibinfo {volume} {121}},\ \bibinfo {pages} {4874} (\bibinfo {year}
  {2004})\BibitemShut{NoStop}%
\bibitem{Palermo}%
  \BibitemOpen
  \bibfield{author}{%
  \bibinfo {author} {\bibfnamefont{V.}~\bibnamefont{Palermo}}, \bibinfo
  {author} {\bibfnamefont{F.}~\bibnamefont{Biscarini}},\ and\ \bibinfo {author}
  {\bibfnamefont{C.}~\bibnamefont{Zannoni}},\ }%
  \bibfield{journal}{%
  \bibinfo {journal} {Phys. Rev. E}\ }%
  \textbf{\bibinfo {volume} {57}},\ \bibinfo {pages} {2519} (\bibinfo {year}
  {1998})\BibitemShut{NoStop}%
\bibitem{Garnier}%
  \BibitemOpen
  \bibfield{author}{%
  \bibinfo {author} {\bibfnamefont{F.}~\bibnamefont{Garnier}}, \bibinfo
  {author} {\bibfnamefont{A.}~\bibnamefont{Yassar}}, \bibinfo {author}
  {\bibfnamefont{R.}~\bibnamefont{Hajlaoui}}, \bibinfo {author}
  {\bibfnamefont{G.}~\bibnamefont{Horowitz}}, \bibinfo {author}
  {\bibfnamefont{F.}~\bibnamefont{Deloffre}}, \bibinfo {author}
  {\bibfnamefont{B.}~\bibnamefont{Servet}}, \bibinfo {author}
  {\bibfnamefont{S.}~\bibnamefont{Ries}},\ and\ \bibinfo {author}
  {\bibfnamefont{P.}~\bibnamefont{Alnot}},\ }%
  \bibfield{journal}{%
  \bibinfo {journal} {J. Am. Chem. Soc.}\ }%
  \textbf{\bibinfo {volume} {115}},\ \bibinfo {pages} {8716} (\bibinfo {year}
  {1993})\BibitemShut{NoStop}%
\bibitem{Bai}%
  \BibitemOpen
  \bibfield{author}{%
  \bibinfo {author} {\bibfnamefont{Q.}~\bibnamefont{Zeng}}, \bibinfo {author}
  {\bibfnamefont{C.}~\bibnamefont{Wang}}, \bibinfo {author}
  {\bibfnamefont{B.}~\bibnamefont{Zhang}}, \bibinfo {author}
  {\bibfnamefont{S.}~\bibnamefont{Xu}}, \bibinfo {author}
  {\bibfnamefont{P.}~\bibnamefont{Wu}}, \bibinfo {author}
  {\bibfnamefont{X.}~\bibnamefont{Qiu}},\ and\ \bibinfo {author}
  {\bibfnamefont{C.}~\bibnamefont{Bai}},\ }%
  \bibfield{journal}{%
  \bibinfo {journal} {J. Indian Chem. Soc.}\ }%
  \textbf{\bibinfo {volume} {77}},\ \bibinfo {pages} {599} (\bibinfo {year}
  {2000})\BibitemShut{NoStop}%
\bibitem{Ediger}%
  \BibitemOpen
  \bibfield{author}{%
  \bibinfo {author} {\bibfnamefont{K.~L.}\ \bibnamefont{Kearns}}, \bibinfo
  {author} {\bibfnamefont{S.~F.}\ \bibnamefont{Swallen}}, \bibinfo {author}
  {\bibfnamefont{M.~D.}\ \bibnamefont{Ediger}}, \bibinfo {author}
  {\bibfnamefont{T.}~\bibnamefont{Wu}}, \bibinfo {author}
  {\bibfnamefont{Y.}~\bibnamefont{Sun}},\ and\ \bibinfo {author}
  {\bibfnamefont{L.}~\bibnamefont{Yu}},\ }%
  \bibfield{journal}{%
  \bibinfo {journal} {J. Phys. Chem. B}\ }%
  \textbf{\bibinfo {volume} {112}},\ \bibinfo {pages} {4934} (\bibinfo {year}
  {2008})\BibitemShut{NoStop}%
\bibitem{Pastor_1}%
  \BibitemOpen
  \bibfield{author}{%
  \bibinfo {author} {\bibfnamefont{F.}~\bibnamefont{Rom\'a}}, \bibinfo {author}
  {\bibfnamefont{A.~J.}\ \bibnamefont{Ramirez-Pastor}},\ and\ \bibinfo {author}
  {\bibfnamefont{J.~L.}\ \bibnamefont{Riccardo}},\ }%
  \bibfield{journal}{%
  \bibinfo {journal} {Phys. Rev. B}\ }%
  \textbf{\bibinfo {volume} {68}},\ \bibinfo {pages} {205407} (\bibinfo {year}
  {2003})\BibitemShut{NoStop}%
\bibitem{Pastor_2}%
  \BibitemOpen
  \bibfield{author}{%
  \bibinfo {author} {\bibfnamefont{F.}~\bibnamefont{Rom\'a}}, \bibinfo {author}
  {\bibfnamefont{A.~J.}\ \bibnamefont{Ramirez-Pastor}},\ and\ \bibinfo {author}
  {\bibfnamefont{J.~L.}\ \bibnamefont{Riccardo}},\ }%
  \bibfield{journal}{%
  \bibinfo {journal} {Phys. Rev. B}\ }%
  \textbf{\bibinfo {volume} {72}},\ \bibinfo {pages} {035444} (\bibinfo {year}
  {2005})\BibitemShut{NoStop}%
\bibitem{Rzysko}%
  \BibitemOpen
  \bibfield{author}{%
  \bibinfo {author} {\bibfnamefont{W.}~\bibnamefont{R\.{z}ysko}}\ and\ \bibinfo
  {author} {\bibfnamefont{M.}~\bibnamefont{Bor\'{o}wko}},\ }%
  \bibfield{journal}{%
  \bibinfo {journal} {J. Chem. Phys.}\ }%
  \textbf{\bibinfo {volume} {135}},\ \bibinfo {pages} {194702} (\bibinfo {year}
  {2011})\BibitemShut{NoStop}%
\bibitem{Pablo}%
  \BibitemOpen
  \bibfield{author}{%
  \bibinfo {author} {\bibfnamefont{S.}~\bibnamefont{Sastry}}, \bibinfo {author}
  {\bibfnamefont{P.~G.}\ \bibnamefont{Debenedetti}},\ and\ \bibinfo {author}
  {\bibfnamefont{F.~H.}\ \bibnamefont{Stillinger}},\ }%
  \bibfield{journal}{%
  \bibinfo {journal} {Nature}\ }%
  \textbf{\bibinfo {volume} {393}},\ \bibinfo {pages} {554} (\bibinfo {year}
  {1998})\BibitemShut{NoStop}%
\bibitem{Kirkpatrick1983}%
  \BibitemOpen
  \bibfield{author}{%
  \bibinfo {author} {\bibfnamefont{S.}~\bibnamefont{Kirkpatrick}}, \bibinfo
  {author} {\bibfnamefont{C.~D.}\ \bibnamefont{Gelatt~Jr.}},\ and\ \bibinfo
  {author} {\bibfnamefont{M.~P.}\ \bibnamefont{Vecchi}},\ }%
  \bibfield{journal}{%
  \bibinfo {journal} {Science}\ }%
  \textbf{\bibinfo {volume} {220}},\ \bibinfo {pages} {671} (\bibinfo {year}
  {1983})\BibitemShut{NoStop}%
\end{thebibliography}
%

\end{document}